\documentstyle[12pt]{article}
\newcommand{\bolddot}[0]{\mbox{$\cdot$}}
\newcommand{\zop}[0]{Z_{0+}}
\newcommand{\zex}[0]{Z^{ext}_N}
\newcommand{\zin}[0]{Z^{int}_N}
\newcommand{\gex}[0]{\Gamma^{ext}}
\newcommand{\gin}[0]{\Gamma^{int}}
\newcommand{\nal}[0]{n_{\alpha}}
\newcommand{\mdot}[0]{m_{\bolddot}}
\newcommand{\ndot}[0]{n_{\bolddot}}
\newcommand{\undot}[0]{\ul{n}_{\bolddot}}
\newcommand{\xdot}[0]{x_{\bolddot}}

\newcommand{\edot}[0]{E_{\bolddot}}
\newcommand{\hdot}[0]{H_{\bolddot}}
\newcommand{\calh}[0]{{\cal H}}

\newcommand{\bra}[1]{\langle#1|}  
\newcommand{\ket}[1]{|#1\rangle}
\newcommand{\av}[1]{\langle#1\rangle}
\newcommand{\pder}[2]{\frac{\partial#1}{\partial#2}}

\newcommand{\beq}{\begin{equation}}  
\newcommand{\eeq}{\end{equation}}
\newcommand{\beqa}{\begin{eqnarray*}}  
\newcommand{\eeqa}{\end{eqnarray*}}
\newcommand{\rarrow}[0]{\rightarrow}
\newcommand{\ul}[1]{\underline{#1}}
\newcommand{\ol}[1]{\overline{#1}}
\newcommand{\enote}[1]{\cite{#1}}    
\newcommand{\eqlabel}[1]{\renewcommand{\theequation}{#1}}

\begin{document}
\title{QUANTUM BAYESIAN NETS}

\author{Robert R. Tucci\\
        P.O. Box 226,\\
        Bedford, MA 01730}

\date{Published in: Int. Jour. Of Mod. Phys. B9(1995)295-337\\
The principles of this paper have been implemented\\
 in commercial software available at\\
www.ar-tiste.com} 

\maketitle

\vskip2cm
\section*{ABSTRACT}We begin with a review of a well known class of
networks, Classical Bayesian (CB) nets (also called 
causal probabilistic nets by some). Given a situation which includes
randomness, CB nets are used to calculate the probabilities of various
hypotheses about the situation, conditioned on the available evidence. We
introduce a new class of networks, which we call Quantum Bayesian (QB)
nets, that generalize CB nets to the quantum mechanical regime. We
 explain how to use QB nets  to 
calculate quantum mechanical conditional probabilities (in case of either
sharp or fuzzy observations), and discuss the connection of QB nets to
Feynman Path integrals. We give  examples  of QB nets 
 that involve
 a single spin-$\frac{1}{2}$ particle passing through a
 configuration of two or three Stern-Gerlach magnets. For the examples
given, we present the numerical values of various conditional
probabilities,  as calculated by a general computer program
especially written for
this purpose.  

\newpage

\section*{1. INTRODUCTION}

The  artificial intelligence and expert systems literature contains a
large number of articles (\enote{Pop1}-\enote{Pearl})  discussing the theory and
application of Classical Bayesian (CB) nets. At least one software
package\enote{ERGO} that implements this theory is commercially  available.
 Amazingly, the physics community, to whom this paper is mainly
addressed,  seems not to have discovered or used CB nets yet.
Therefore, we  begin this paper
with a  review of  CB nets. However, the
real purpose of this paper is to 
 introduce a new class of nets, which we
shall call Quantum Bayesian (QB) nets, that 
 generalize  CB nets to the quantum mechanical regime. 
In this paper, to illustrate  QB nets,  we use them to
predict  conditional probabilities for experiments comprising
combinations of two or three Stern-Gerlach magnets. In future
papers, we will use QB nets to analyze quantum optical experiments and
other physical situations. Earlier workers have devised networks for
quantum mechanics ( for eg., Feynman  diagrams, and the ``trajectory
graphs" of Ref.\enote{Griffiths}), but their nets differ substantially
from ours.

Henceforth, we will underline random variables. 
We will write $P(\ul{x}=x)$ for the probability that the
random variable $\ul{x}$ assumes the particular value
$x$\enote{more precisely}. Sometimes, if there is no danger of 
confusion, we will write $P(x)$ rather than  $P(\ul{x}=x)$. Similarly, we
will often write $P(y | x)$ instead of   $P(\ul{y}=y | \ul{x}=x)$. 
$P(\ul{y}=y | \ul{x}=x)$, the conditional probability that $\ul{y}$
assumes the value $y$ given that $\ul{x}$ assumes the value $x$, is
defined by 

\beq
P(\ul{y}=y | \ul{x}=x)= 
\frac{ P(\ul{y}=y, \ul{x}=x)}
{P(\ul{x}=x)}
\;.
\eqlabel{1.1}\eeq

Suppose $\ul{h}$ is a random variable that can take on
as values each of a number of {\it hypotheses}, 
 and $\ul{e}$ is a random variable representing 
the observations  that constitute the available 
{\it evidence}. 
We often want to calculate the posterior probability 
$P(h | e)$ in terms of the
prior probability $P(h)$.   To do 
this, one may use Bayes's rule,  which
says

\beq
P(h |e)= \frac{P(e|h)P(h)}
{\sum_{h'} P(e|h') P(h')}
\;.
\eqlabel{1.2}\eeq
Equation(1.2) is a  simple consequence of
Eq.(1.1).
The subject of CB nets may be viewed as an  extension  of Bayes's rule.

A CB net has  two parts: a diagram consisting of nodes with
arrows connecting some pairs of these nodes, and a collection of
probabilities,  one per node. For example, the
digital circuits of  electrical engineering can be modelled as CB nets
if each NAND (or, alternatively, each AND, OR and NOT gate) is
replaced by a node, and each connecting cable is replaced by an arrow 
pointing in the direction of current flow. Of course, in the usual
digital circuits, the NAND gates are deterministic, their output
being a deterministic function of their inputs. More generally, if
some of the NAND gates were to act erratically, and if the input
sources to the circuit also acted in a random fashion, then 
such a circuit could also be modelled by
a CB net. In this non-deterministic CB net, the signal flowing in each
wire would have a certain probability of being 0 and of being 1. If the
signal in a certain wire were measured so that one knew that it was
definitely 1, then one would want to revise the probability distributions
for the signals in all other wires so as to reflect the new evidence.
Hence, conditional probabilities and Bayes's rule arise naturally when
considering probabilistic nets.

As we shall see, a major difference between CB nets and QB nets is
that for QB nets, one assigns complex amplitudes rather than
probabilities to each node.

\section*{2. THEORY OF CB NETS}
In this section, we will review the simple theory of CB nets. The
next section will present some examples.

We call a {\it graph} (or a diagram or an
architecture) a collection of nodes with arrows connecting some pairs
of these nodes. The arrows of the graph must satisfy certain
constraints that will be specified below. We call a {\it labelled graph}
a graph whose nodes are
labelled.  A
{\it CB net}  consists of two parts:  a  labelled graph with each node
labelled by a random variable,  and   a collection of node matrices,
one matrix for each node. These two parts must satisfy certain
constraints that will be specified below. 

We define two kinds of arrows: {\it internal arrows}
are those that have  a starting node and a different ending one;
{\it external arrows} are those that have a starting node but no ending
one.  We  define two
types of nodes:  {\it external nodes} are  those that  have  a single
external arrow leaving it, and {\it internal nodes} are  those that have
one or more internal arrows leaving it. It is also common to
use the terms {\it root node} or {\it prior probability node} for a node
which has no incoming arrows, only outgoing ones. We restrict all nodes
of a graph to be either internal or external. Hence, no nodes may have
both an external and one or more internal arrows leaving it. 

We define each
node of a CB net to represent a numerical value, and the  whole net  to
represent the product of all these node values. We assign a numerical
value to each node as follows. First, we assign a
 random variable to each node. Suppose the random
variables assigned to the $N$ nodes are $\ul{x}_1,\ul{x}_2,\cdots
,\ul{x}_N$. Define $Z_N=\{1,2,\cdots,N\}$. 
For any finite set $S$, let $|S|$ be the number of elements in $S$. 
If $S=\{k_1,k_2,\cdots, k_{|S|}\}\subset Z_N$,
define $(x_{\bolddot})_S=(x_{k_1},x_{k_2},\cdots,x_{k_{|S|}})$ and 
$(\ul{x}_{\bolddot})_S=(\ul{x}_{k_1},\ul{x}_{k_2},\cdots,\ul{x}_{k_{|S|}})$.
Sometimes, we also abbreviate $(x_{\bolddot})_{Z_N}$ 
(i.e., the vector that includes all the possible $x_j$ components) by just
$x_{\bolddot}$, and $(\ul{x}_{\bolddot})_{Z_N}$ by just 
$\ul{x}_{\bolddot}\;$. 
For $j\in Z_N$, we imagine node $\ul{x}_j$ to lie in state 
 $x_j$ (See Fig.1).  
 We also
imagine all arrows leaving the node $\ul{x}_j$ to lie
 in state $x_j$, and thus
we label all of them $x_j$. At this point we've
shown how to label each arrow in the graph by $x_k$ for some $k\in Z_N$.
Define $S_j$ to be the set of all $k$ such that 
an arrow labelled $x_k$ enters node 
$\ul{x}_j$. 
   Now we
assign a value $P[x_j|(x_{\bolddot})_{S_j}]$ to node $\ul{x}_j$. 
$P[x_j|(x_{\bolddot})_{S_j}]$ is
what we referred to earlier as a {\it node matrix}; $x_j$ is the matrix's
{\it row index} and   $(x_{\bolddot})_{S_j}$ is
its {\it column index}.  As the notation suggests, we assume that the
values  $P[x_j|(x_{\bolddot})_{S_j}]$ are conditional probabilities;
i.e., that they satisfy

\beq
 P[x_j|(x_{\bolddot})_{S_j}] \geq 0
\;,
\eqlabel{2.1}\eeq 

\beq
\sum_{x_j} P[x_j|(x_{\bolddot})_{S_j}] =1
\;,
\eqlabel{2.2}\eeq 
where the sum in Eq.(2.2) is over all states $x_j$ that the
random variable $\ul{x}_j$ can assume, and where Eqs. (2.1) and (2.2) must
be satisfied for all $j\in Z_N$ and for all possible values of the 
vector $(\ul{x}_{\bolddot})_{S_j}$ of random variables.  
The left hand side of Eq.(2.2) is just the sum over the entries  along
a column of a node matrix.
The CB net is
taken to represent the product of all the probabilities 
$P[x_j|(x_{\bolddot})_{S_j}]$ for  $j\in Z_N$. This product is a function
$P(\xdot)$ of the current states $x_1, x_2,\cdots,x_N$ of
the nodes. Thus,

\beq
P(\xdot)=\prod_{j\in Z_N}
P[x_j|(x_{\bolddot})_{S_j}]
\;.
\eqlabel{2.3}\eeq 
We require that

\beq
\sum_{\xdot} P(\xdot) =1
\;,
\eqlabel{2.4}\eeq
as expected for a joint probability distribution of the random variables 
$\ul{x}_1, \ul{x}_2,\cdots,\ul{x}_N$.

Next, to illustrate the CB net concepts just presented,  we
will discuss all possible CB nets with two or three nodes.

	For 2 nodes labelled by random variables $\ul{x}$ and 
$\ul{y}$, there 
are only 2 possible labelled graphs, depending on whether the arrow
points  from $\ul{x}$ to $\ul{y}$ (Fig.2a) or vice versa (Fig.2b).

	Figure 2a is diagrammatic notation for the following 
equation:

\beq	
P(x,y)=P(y|x)P(x)
\;.
\eqlabel{2.5}\eeq
Notice that each probability factor on the right hand side of 
Eq.(2.5) is represented by a node in Fig.2a.  Notice also that 
the prior probability $P(x)$, since it has no conditions placed on it, is 
portrayed in Fig.2a by a 
 node which has no arrows 
pointing into it.  The probability $P(y|x)$, on the other hand,
depends on  the value $x$ of node $\ul{x}$, and thus it is represented
by a node with an  arrow labelled $x$ pointing into it. According to
Eq.(1.1),  Eq.(2.5) is a {\it tautology} (i.e., a statement that
is always true). Thus, the net of Fig.2a  may represent any probability
distribution of two random variables $\ul{x}$ and $\ul{y}$.

Figure 2b is diagrammatic notation for the following equation:

\beq
P(x,y)=P(x|y)P(y)
\;.
\eqlabel{2.6}\eeq
Again, this equation is a tautology.  Thus, like the net of Figs.2a,
the net of Fig.2b may represent an arbitrary
probability  distribution of two random variables $\ul{x}$ and $\ul{y}$.
(A collorary is that two  CB
nets with different labelled graphs 
may still represent the same probability distribution. )

	Figure 3a is diagrammatic notation for

\beq
P(x,y,z)=P(z|x,y)P(y)P(x)
\;.
\eqlabel{2.7}\eeq
Contrary to Eqs.(2.5) and (2.6), Eq.(2.7) does not represent a 
tautology:  not all probability distributions of three random
variables  $\ul{x}$, $\ul{y}$ and $\ul{z}$ must satisfy
Eq.(2.7). In fact, 
summing  both sides of the last equation over the values of $\ul{z}$
yields

\beq
P(x,y)=P(y)P(x)
\;,
\eqlabel{2.8}\eeq
i.e., $\ul{x}$ and $\ul{y}$ are {\it independent}.  
Thus, the net of Fig.3a 
represents two independent random variables $\ul{x}$ and $\ul{y}$.

	Figure 3b is diagrammatic notation for the following 
equation:

\beq
P(x,y,z)=P(z|x)P(y|x)P(x)
\;.
\eqlabel{2.9}\eeq
Dividing both sides of  Eq.(2.9) by $P(x)$
yields:

\beq
P(y,z|x)=P(z|x)P(y|x)
\;.
\eqlabel{2.10}\eeq
Even though $\ul{y}$ and $\ul{z}$ are not necessarily independent, 
they are  {\it conditionally independent}, i.e., they are independent
 at fixed
$\ul{x}=x$.   Thus, the net of Fig.3b represents two
conditionally  independent random variables $\ul{y}$ and $\ul{z}$.

Figure 3c is diagrammatic notation for the equation:

\beq
P(x,y,z)=P(z|y)P(y|x)P(x)
\;.
\eqlabel{2.11}\eeq
Random variables $\ul{x}_1,\ul{x}_2,\cdots,\ul{x}_N$ form an 
{\it N step Markov chain} ($N$ may be infinite) if
$P(x_{n+1}|x_n,x_{n-1},\cdots,x_1)=
P(x_{n+1}|x_n)$ for $n=1,2,\cdots,N-1$, i.e., if the (n+1)th step 
$x_{n+1}$
depends only on the step $x_n$ immediately preceding it. Clearly, the random
variables  $\ul{x}$, 
$\ul{y}$
and  $\ul{z}$ of Fig.3c represent a 3 step Markov chain.

Figure 3d is 
diagrammatic notation for the equation:

\beq
P(x,y,z)=P(z|y,x)P(y|x)P(x)
\;.
\eqlabel{2.12}\eeq
Replacing the conditional probabilities on the right hand side of 
Eq.(2.12) by their definitions in terms of probabilities without
conditions, we obtain $P(x,y,z)$  for the right hand side. Thus,
Eq.(2.12)  is a tautology, and the net of Fig.3d can represent 
any probability distribution of three random variables $\ul{x}$,
$\ul{y}$ and $\ul{z}$.

	The graph of Fig.3d is {\it acyclic}.  That is, it does not
contain  any cycles (a {\it cycle} is a closed path of arrows with the arrows all
pointing in the same sense).
 On the other hand, the
graph of Fig.4  is {\it paracyclic} (i.e., it contains at least one
cycle). In fact, the whole graph of Fig.4 is a cycle. The net 
of  Fig.4 shall be forbidden  because, if one sums over its free indices
($x,y,z$), one does not always obtain unity, as must be the case if the
net is to represent a probability distribution $P(x,y,z)$. For example,
if we assume that $\ul{x},\ul{y}$ and $\ul{z}$ can assume only states 0
and 1, and if 
 $P(y|x)=\delta_{y,x}$,
$P(x|z)=\delta_{x,z}$ and $P(z|y)=\delta_{z,y}$, where 
$\delta_{x,y}$ is the Kronecker delta function, then  Fig.4 adds up
to $\sum_{x,y,z}\delta_{x,z}\delta_{z,y}\delta_{y,x}=2$.
The acyclic net of Fig.3d does
add up to one, as proven in diagrammatic notation in Fig.5. 
In this figure, summation
over the states of an arrow is indicated by giving the arrow a double
shaft. Note that in
Fig.5, we first add over the index $z$ of the  external arrow.  This
produces a new external arrow, and we add  over its index $y$,  and so
on.  Each time we add over the index of the  current external arrow until
finally we get unity.  One cannot follow  this summation procedure to
show that the net of Fig.4 adds up to one.   Indeed, Fig.4 has no
external arrow to begin the  procedure with.

Certain aspects of the preceding discussion of  nets  with two or three 
nodes can be generalized to any number $N$ of nodes. 

For any number of nodes, we will say that a graph is {\it fully
connected} if it is acyclic, and if everyone of its nodes is connected
to all other nodes by either incoming or outgoing arrows. We will say
that a net is fully connected if its graph is fully connected.
	Any fully connected CB net represents a 
 completely general joint probability  distribution of the
random variables labelling its nodes. Indeed, it is always possible
to label the nodes of an $N$ node fully connected CB net so that the
net represents the right hand side of the following equation:

\beq
 P(\xdot)=
P(x_N|x_{N-1},x_{N-2},\cdots,x_1)
P(x_{N-1}|x_{N-2},x_{N-3}\cdots,x_1)
\cdots
P(x_{2}|x_1) P(x_1)
\;.\eqlabel{2.13}\eeq
And this last equation is a tautology. To label the nodes of a fully
connected net so that Eq.(2.13) applies to it, one proceeds as follows.
There always exists exactly one external node. This is why. The
external node, call it $x_N$, must exist. Otherwise, all nodes would have
at least one outgoing internal arrow. Then one could start from any node
and travel along one of its outgoing internal arrows to reach another
node, and so on, until one came back to a previously visited node.
Therefore, the graph would not be acyclic. The external node $x_N$ is
unique because, since all other nodes have outgoing arrows that point to
$x_N$,  all other nodes must be internal. Remove $x_N$ from the graph.
For the same reasons as before, the  resulting diminished graph contains
a unique external node. Call the latter node $x_{N-1}$.  Continue
removing nodes in this way until all the nodes are labelled
$x_1,x_2,\cdots,x_N$. Since Eq.(2.13) suggests that $x_j$ occurs after $x_{j-1}$ for
$j= 2,3,\cdots,N$, we call this  
node labelling (ordering)  the
{\it chronological labelling} of
the graph.  Two colloraries of the preceding proof are that fully
connected graphs have a single external node, and that all fully
connected  $N$-node labelled graphs are identical once they are relabelled
in the fashion described above. Figure 6a shows the fully connected four
node graph with its chronological labelling. 
By deforming Fig.6a into a topologically equivalent diagram, one
obtains Fig.6b, a more ``stylized'' version of the same thing.
Fig.6b might make more clear to some readers how the arrows of a
fully-connected graph are organized.

Note that one can relabel any graph chronologically, even if it isn't
fully connected. Indeed, given any graph $G$, one may add arrows to it
to form a fully connected graph $\ol{G}$. Call $\ol{G}$ a
{\it completion} of $G$. Sometimes it is possible to add arrows to $G$
in two different ways and arrive at two different completions. Hence
completions are not always unique.  Any completion of a graph $G$ may be
labelled chronologically following the procedure described above. This, in
turn, gives the graph $G$ a chronological ordering, albeit, not a
necessarily unique one. Suppose that $\ul{x}$ and $\ul{y}$ are two nodes
in a graph $G$. We  say that $\ul{x}$ {\it precedes} $\ul{y}$ and write
$\ul{x} < \ul{y}$  if, for any completion of $G$, there exist integers
$j$ and $j'$ with $j<j'$  so that $\ul{x}=\ul{x}_j$ and
$\ul{y}=\ul{x}_{j'}$. We say that $\ul{x}$ and $\ul{y}$ are {\it
concurrent} and write  $\ul{x}\sim \ul{y}$  if  $\ul{x}$ precedes
$\ul{y}$ in some completions but $\ul{y}$ precedes $\ul{x}$ in others.
And we say that  $\ul{x}$ {\it succeeds} or {\it follows} $\ul{y}$ and
write  $\ul{x} > \ul{y}$ if 
 $\ul{y}< \ul{x}$.

Call a {\it CB  pre-net} a labelled graph and an
accompanying set of node matrices that satisfy Eqs.(2.1), (2.2) and
(2.3), but don't necessarily  satisfy the overall
normalization condition Eq.(2.4). An acyclic  CB pre-net
always satisfies Eq.(2.4), and a paracyclic CB pre-net may not
satisfy Eq.(2.4). This is why. Following the procedure just discussed, the
nodes of any acyclic pre-net can be relabelled chronologically as
$\ul{x}_1,\ul{x}_2,\cdots,\ul{x_N}$.  The  relabelled  graph, even if it
is not fully connected, corresponds to the right hand side of Eq.(2.13),
except that some of the conditional probabilities on the right hand side
of Eq.(2.13) might include redundant conditioning. (If a conditional
probability   $P(x|y,R)$ is known to satisfy $P(x|y,R)=P(x|y)$, then we
would say that the expression $P(x|y,R)$ shows redundant conditioning on
$R$.)  This correspondence between any acyclic pre-net and the right hand
side of Eq.(2.13)  guarantees that acyclic pre-nets will satisfy the
overall normalization condition Eq.(2.4). 
On the other hand, as we've shown with the 3-node
pre-net of Fig.4, an $N$-node  paracyclic pre-net
  may not  reduce to unity upon
summing over its free indices. If one tries the process of
``adding over all the current external arrows" on 
a pre-net with a cycle embedded in
it, at some point the process comes to a stop and cannot be completed due
to a lack of a current external arrow to sum over next. 
Note that, in  some sense, Eq.(2.13) embodies
the {\it principle of causality}. Thus, acyclic CB pre-nets preserve 
causality, and
paracyclic CB pre-nets violate this principle.  
If one considers only acyclic
graphs, as we shall do henceforth, then 
there is no difference between CB nets and CB pre-nets. 

Note that if one sums  both sides of Eq.(2.11) over $y$ and divides by
$P(x)$, one obtains

\beq
P(z|x)=\sum_y P(z|y) P(y|x)
\;.
\eqlabel{2.14}\eeq
This last equation, valid for a Markov chain $(\ul{x},\ul{y},\ul{z})$,
is the so called Chapman-Kolgomorov equation. Equation (2.14)
is represented diagramatically in Fig.7.  If one sums  
both sides of Eq.(2.12) over $y$ and divides by
$P(x)$, one obtains

\beq
P(z|x)=\sum_y P(z|y,x) P(y|x)
\;.
\eqlabel{2.15}\eeq
 The last equation is a generalization of the
Chapman-Kolgomorov equation to arbitrary random variables 
$\ul{x},\ul{y},\ul{z}$ that don't necessarily form a Markov chain.
Equation (2.15)
is represented diagramatically in Fig.8.
Note that in both Figs.7 and 8, we follow a process of adding over some
of the  arrows of a net to obtain a new net with fewer nodes.
This process may be called {\it coarsening} or {\it data compression},
because it reduces the number of nodes and because the joint
probability distribution of the new net carries less information than the
joint probability distribution of the old one. In general, if we start
with $N$ nodes  $\ul{x}_1, \ul{x}_2,\cdots, \ul{x}_N$, and we sum over 
 $x_1,x_2,\cdots,x_k$, then the resulting probability distribution of the
variables  $x_{k+1},x_{k+2},\cdots,x_N$  can always be represented by a
fully connected net with nodes 
$\ul{x}_{k+1},\ul{x}_{k+2},\cdots,\ul{x}_N$.

In  modelling a classical physical
situation by a CB net, if one knows very little about the nature of the
probability distributions involved, one can  always use a fully
connected net.  Later on, if one learns that certain  pairs of random
variables are conditionally independent, one may remove the
arrows connecting these pairs of variables without
changing the value of the full net.

Once we have designed  a net architecture and the net's node 
matrices have been
calculated, what next? How to use this information?
 The information is sufficient for calculating  the joint probability 
$P(x_1,x_2,\cdots,x_N)$, where 
$\ul{x}_1,\ul{x}_2,\cdots,\ul{x}_N$ are  the nodes of the net.
From this joint probability, one can 
calculate $P(x_a|x_b,x_c,\cdots)$, the probability that  any
one node $\ul{x}_a$ of the net assumes one of its states $x_a$,
conditioned on several  other nodes  $\ul{x}_b,\ul{x}_c,\cdots$ of
the net assuming respective states
$x_b,x_c,\cdots$. To go from $P(x_1,x_2,\cdots,x_N)$ to
$P(x_a|x_b,x_c,\cdots)$, one can  sum $P(x_1,x_2,\cdots,x_N)$ over those 
variables which are not in the list $x_a,x_b,x_c,\cdots$ to get 
$P(x_a,x_b,x_c,\cdots)$.
 One can obtain $P(x_b,x_c,\cdots)$ similarly, and
then divide the two. However, this brute force method takes no advantage of the
particular topology of the net,  and so it is very labor intensive. 
Artificial intelligence and expert systems  workers have found
an algorithm for calculating  
 $P(x_a|x_b,x_c,\cdots)$ from $P(x_1,x_2,\cdots,x_N)$
that takes advantage of the net topology to reduce dramatically the
number of arithmetical operations required. For a discussion of
this great achievement, the reader can consult 
Refs.\enote{Pop1}-\enote{Pearl}.
The software package of Ref.\enote{ERGO} implements 
  this fast algorithm. 

\section*{3.  EXAMPLES OF CB NETS}

The references at the end of this paper present examples of CB nets used
in medical diagnosis\enote{LS},\enote{IBM}, monitoring of
processes\enote{Pop1},  genetics\enote{Munin},
etc.
And this is just a small fraction of the possible applications of CB
nets. Indeed, any probabilistic model can be discussed in terms of CB
nets. In this section, we do not intend to present a
comprehensive  collection of CB net examples, but only to give a few
simple introductory ones.

(a){\sc digital circuits}

Figure 9 shows a Bayesian net version of an AND gate. The random
variables $\ul{x}, \ul{y}$ and $\ul{z}$ are {\it binary}
 (i.e. they assume
values in $\{0,1\}$.) The node matrices for this net consist of the
prior probabilities $P(x), P(y)$ and of the conditional probability 
$P(z|x,y)$. $P(z|x,y)$ is given by the table in Fig.9.
With states 0 and 1 standing for false and true, respectively, 
 this table is what one would expect if  $z=$  ( $x$ and
$y$ ). Note that the columns of the node matrix  $P(z|x,y)$ add up to
one, as required by Eq.(2.2). A node matrix whose entries are all
either 0 or 1 (like $P(z|x,y)$ in this example) will be  called a {\it
deterministic node matrix}.

CB nets representing OR and NOT gates can be defined analogously
 to the AND
net. Then one can construct CB nets for any combination of AND, OR and NOT
gates.

(b){\sc constraint nodes}

Figure 10 shows a net for the binary random variables $\ul{x}, \ul{y}$
and the random variable $\ul{z}\in\{0,1,2\}$. To fully specify this
net, one must give the prior probabilities $P(x), P(y)$ and the
conditional probability $P(z|x,y)$. The node matrix
$P(z|x,y)$ is given by the table of Fig.10. This node matrix is
deterministic and agrees with $z=x+y$. Note that if we know for
sure that $x+y=1$, then we can ``fix" the {\it sum node} $\ul{z}$
to one, and calculate probabilities, like $P(x|\ul{z}=1)$, which
have  $\ul{z}=1$ as evidence. In such a case, we call $\ul{z}$ a {\it
constraint node}, because it is used to enforce the constraint
$x+y=1$.

As another example of a net that possesses a node which is useful to
constrain, consider Fig.11. In this net the random variables
$\ul{x},\ul{y}$ and $\ul{z}$ are binary. To fully specify the net, one
must give prior probabilities $P(x)$, $P(y)$ and the conditional
probability $P(z|x,y)$. $P(z|x,y)$ is given by the table in Fig.11.
This table is  what one would expect if 
 $z=$ (if $x$ then $y$). Since an ``if $x$  then $y$" statement does
not say anything about what to do when $x=0$, the probabilities
$P(z|x,y)$ are arbitrary and not necessarily deterministic
 when $x=0$. The
net of Fig.11 could be used by fixing the {\it if-then node} $\ul{z}$ to
one and considering only probabilities with $\ul{z}=1$ as evidence. In
such a case, we  would say that the $\ul{z}$ node was a constraint node.

(c){\sc clauser-horne experiment}

Consider the Clauser-Horne experiment, which is used to observe
violations of Bell-type inequalities\enote{Tucci}.  Two
particles, call them 1 and 2, are created at a common vertex and fly
apart. Let $\lambda$ represent the ``hidden variables". Particle 1 is
subjected to a spin measurement 
along the direction $A$  with outcome $x^A_1\in\{+,-\}$ or 
along the direction $A'$  with outcome $x^{A'}_1\in\{+,-\}$. 
 Particle 2 is subjected to a spin measurement 
along the direction $B$  with outcome $x^B_2\in\{+,-\}$ or 
along the direction $B'$  with outcome $x^{B'}_2\in\{+,-\}$. 
One can draw the net
of Fig.12, which has nodes $\ul{\lambda},\ul{x}^{\theta_1}_1$
and  $\ul{x}^{\theta_2}_2$. (Figure 12 really represents 4 nets, one for
each of the following possibilities:  
$(\theta_1, \theta_2) =(A,B), (A,B'), (A',B), (A',B')$). To specify
the net, one must give  probabilities $P(\lambda)$,
$P(x^{\theta_1}_1|\lambda)$ and  $P(x^{\theta_2}_2|\lambda)$. Note that
if one sums the net of Fig.12 over $\lambda$, one gets

\beq
P(x^{\theta_1}_1,x^{\theta_2}_2)=
\sum_{\lambda}
P(x^{\theta_1}_1|\lambda) P(x^{\theta_2}_2|\lambda)
P(\lambda)
\;,
\eqlabel{3.1}\eeq
which is the starting point in the derivation of the Bell inequalities
for the Clauser-Horne experiment. 

As a slightly more complicated experiment, one might choose at random
whether $\theta_1=A$ or $A'$ and whether $\theta_2=B$ or $B'$. Such an experiment can be represented by a CB net
with nodes $\ul{\theta}_1 , \ul{x}_1, \ul{\theta}_2 , \ul{x}_2$ and
$\ul{\lambda}$, where
$\ul{x}_j$ for $j\in\{1,2\}$ represents the outcome of a measurement
on particle $j$. See Fig.13. This net is specified if we give 
probabilities 
$ P(x_1|\theta_1,\lambda), P(\theta_1), 
P(x_2|\theta_2,\lambda), P(\theta_2)$ and $P(\lambda)$.  Summing the net
over $\lambda$ and dividing by 
$P(\theta_1) P(\theta_2)$ yields an equation analogous to Eq.(3.1):

\beq
P(x_1,x_2|\theta_1,\theta_2)=
\sum_{\lambda}
P(x_1|\theta_1,\lambda) P(x_2|\theta_2,\lambda)
P(\lambda)
\;.
\eqlabel{3.2}\eeq

(d){\sc random walk}

Suppose that a particle moves in a straight line,  taking  unit length
steps either forwards or backwards, with probabilities $p_+$ and $p_-$,
respectively, where $p_+ + p_-=1$. Let $x_j$ be the position of the
particle at time $j$, with $j\in\{0,1,2,\cdots,\}$.
Assume $x_0$, the starting position, is zero. Define 
$\Delta x_j$ by $\Delta x_j = x_j - x_{j-1}$, for $j\in\{1,2,\cdots,\}$.
Figure 14a shows  a CB net that represents the probability distribution 
of $\ul{x}_0, \ul{x}_1,\cdots$ and 
$\ul{\Delta x}_1, \ul{\Delta x}_2,\cdots$. 
Clearly, 
$x_j\in\{0,\pm 1,\pm 2,\cdots,\pm j\}$ and 
$\Delta x_j \in\{\pm 1\}$.
To fully specify the net of Fig.14a, we must give the probability
matrices associated with each node. These matrices are

\beq
\begin{array}{l}
P(\ul{x}_0=0)=1
\;\;,\\
P(\ul{\Delta x}_j \pm 1)= p_{\pm}\;\;\mbox{for  $j=1,2,\cdots$}
\;\;,\\
P(\ul{x}_j=y | \ul{x}_{j-1}=x, \ul{\Delta x}_j \pm 1)=
\delta(y, x\pm 1) \;\;\mbox{for  $j=1,2,\cdots$}
\;\;,
\end{array}
\eqlabel{3.3}\eeq
where $\delta$ is the Kronecker delta function. By summing the net of
Fig.14a over all possible values of the indices $\Delta x_j$ that label
the arrows from the $\ul{\Delta x}_j$ to the $\ul{x}_j$ nodes, one obtains
Fig.14b. For this coarser net, one obtains

\beq
\begin{array}{l}
P(\ul{x}_0=0)=1
\;\;,\\
P(\ul{x}_j=y | \ul{x}_{j-1}=x)=
p_+\delta(y, x+1) +
p_-\delta(y, x-1) 
 \;\;\mbox{for  $j=1,2,\cdots$}
\;\;.
\end{array}
\eqlabel{3.4}\eeq
One may go even further: For any $k\in \{0,1,2\cdots\}$, 
one may sum Fig.14b over 
all $x_j$ such that $j\not \in\{0,k\}$. One obtains the net of
Fig.14c, with

\beq
P(\ul{x}_0=0)=1
\;,
\eqlabel{3.5a}\eeq

\beq
P(\ul{x}_k=x_k | \ul{x}_0=0)=
\left(
\begin{array}{c}
r+s\\r
\end{array}
\right)
p_+^r p_-^s
\;,
\eqlabel{3.5b}\eeq
where $r-s=x_k$ and $r+s=k$, and the first factor on the right hand side
of Eq.(3.5b) is a combinatorial factor.

\section*{4.  THEORY OF QB NETS}

In this section, we will define QB nets and explain how to use them to
calculate quantum mechanical conditional probabilities.
The next section will give examples of QB nets.

Like a CB net,  a {\it QB net} consists of two parts: a labelled graph
and a collection of node matrices. These two parts must satisfy certain
constraints that will be specified below.

 External and internal arrows,
external and internal nodes, and root nodes are all defined in the same
way for QB nets as for CB nets. All nodes of a QB net must be either
internal or external.

We define each node of a QB net to represent a numerical value, and
the  whole net  to represent the product of all these node values.
We assign a numerical value to each node as follows. First, we assign a
 random variable to each node. Suppose the random
variables assigned to the $N$ nodes are $\ul{x}_1,\ul{x}_2,\cdots
,\ul{x}_N$. Define $Z_N=\{1,2,\cdots,N\}$.
For any finite set $S$, let $|S|$ be the number of elements
in $S$.  If $S=\{k_1,k_2,\cdots, k_{|S|}\}\subset Z_N$,
define $(x_{\bolddot})_S=(x_{k_1},x_{k_2},\cdots,x_{k_{|S|}})$ and 
$(\ul{x}_{\bolddot})_S=(\ul{x}_{k_1},\ul{x}_{k_2},\cdots,\ul{x}_{k_{|S|}})$.
Sometimes, we also abbreviate $(x_{\bolddot})_{Z_N}$ 
(i.e., the vector that includes all the possible $x_j$ components) by just
$x_{\bolddot}$, and $(\ul{x}_{\bolddot})_{Z_N}$ by just $\ul{x}_{\bolddot}\;$. 
For $j\in Z_N$, we imagine node $\ul{x}_j$ to lie in state 
 $x_j$.  
 We also
imagine all arrows leaving the node $\ul{x}_j$ to lie
 in state $x_j$, and thus
we label all of them $x_j$. At this point we've
shown how to label each arrow in the graph by $x_k$ for some $k\in Z_N$.
Define $S_j$ to be the set of all $k$ such that 
an arrow labelled $x_k$ enters node 
$\ul{x}_j$. 
   Now we
assign a complex number $A[x_j|(x_{\bolddot})_{S_j}]$ to node $\ul{x}_j$. 
$A[x_j|(x_{\bolddot})_{S_j}]$ is
the  {\it node matrix} for node $\ul{x}_j$; $x_j$ is the matrix's 
{\it row index} and   $(x_{\bolddot})_{S_j}$ is
its {\it column index}. 
We require that the quantities 
$A[x_j|(x_{\bolddot})_{S_j}]$ be probability amplitudes that satisfy

\beq
\sum_{x_j} \left|A[x_j|(x_{\bolddot})_{S_j}]\right |^2 =1
\;,
\eqlabel{4.1}\eeq 
where the sum in Eq.(4.1) is over all states $x_j$ that the
random variable $\ul{x}_j$ can assume, and where Eq.(4.1) must
be satisfied for all $j\in Z_N$ and for all possible values of the 
vector $(\ul{x}_{\bolddot})_{S_j}$ of random variables.   The QB net is
 taken to
represent the product of all the probability amplitudes 
$A[x_j|(x_{\bolddot})_{S_j}]$ for  $j\in Z_N$. This product is a function 
$A(\xdot)$ of the current states $x_1, x_2,\cdots,x_N$ of the
nodes. Thus,

\beq
A(\xdot)=\prod_{j\in Z_N}
A[x_j|(x_{\bolddot})_{S_j}]
\;.
\eqlabel{4.2}\eeq 
Let $\zex$ be the set of all $j\in Z_N$ such that $\ul{x}_j$ is
an external node,  and  let $\zin$ be the set of those $j\in Z_N$ such
that 
 $\ul{x}_j$  is an internal node. Clearly, $\zex$ and $\zin$ are disjoint
and their union is $Z_N$. 
We require $A(\xdot)$ to satisfy 

\beq
\sum_{(x_{\bolddot})_{\zex}}
\left|\sum_{(x_{\bolddot})_{\zin}} A(\xdot)\right |^2 =1
\;
\eqlabel{4.3}\eeq
and

\beq
\sum_{\xdot} \left|A(\xdot)\right |^2 =1
\;.
\eqlabel{4.4}\eeq

Note that as a consequence of Eqs.(4.1) and (4.4), given any QB
net, one can construct a special CB net by replacing the value
$A[x_j|(x_{\bolddot})_{S_j}]$ of each node by its magnitude squared. 
We call this special CB
net  the {\it parent CB net}
 of the  QB net from which it was
constructed. We call it so because, given  a parent CB net, one
can  replace the  value of each node by its square root times a phase
factor. For a different choice of phase factors, one  generates a
different QB net. Thus, a parent CB net may be used to generate a
whole family of QB nets.

A {\it QB pre-net} is a labelled graph and an accompanying set of
node matrices that satisfy Eqs.(4.1), (4.2) and (4.3),  but  don't
necessarily satisfy Eq.(4.4). A QB pre-net that is acyclic satisfies
Eq.(4.4), because its parent CB pre-net is acyclic and this implies that
Eq.(4.4) is satisfied.  If one considers only acyclic
graphs, as we shall do henceforth, then there is no 
difference between QB
nets  and QB pre-nets.

In the second quantized formulation of quantum mechanics, one speaks of
$M$ {\it modes} represented by $M$ {\it annihilation operators} 
$a_1, a_2, \cdots, a_M$, 
which obey
certain commutation  relations amongst themselves. Define 
$\zop=\{0,1,2,\cdots\}$ and let $n_i\in \zop$ for $i=1$ to $M$. For $M$
modes, one uses quantum states like

\beq
\frac{ a^{\dagger n_1}_1 }
{\sqrt{n_1 !} }
\frac{ a^{\dagger n_2}_2 }
{\sqrt{n_2 !} }
\cdots
\frac{ a^{\dagger n_M}_M }
{\sqrt{n_M !} }
\ket{0}
\;.
\eqlabel{4.5}\eeq
The  state given by Eq.(4.5) is  specified by a vector
$(n_1,n_2,\cdots,n_M)$ of {\it occupation numbers}. For the rest of this
paper, we will consider only QB nets whose states $x_j$ are vectors 
$(n_{j,1},n_{j,2},\cdots,n_{j,K_j})$ of
occupation numbers. Define $\Gamma$ to be the set of all $\alpha$ such
that  $\nal$ is an occupation number
 of the QB net under consideration. For
example,  $\alpha =(j,1)$ in $n_{j,1}$, where $n_{j,1}$ is a component
of  $x_j=(n_{j,1},n_{j,2},\cdots,n_{j,K_j}$). For any set 
$\Gamma'=\{ \alpha_1, \alpha_2, \cdots, \alpha_{|\Gamma'|}\} \subset
\Gamma$,  let 
$(n_{\bolddot})_{\Gamma'}=(n_{\alpha_1},n_{\alpha_2},
\cdots,n_{\alpha_{|\Gamma'|}})$
and 
$(\ul{n}_{\bolddot})_{\Gamma'}=(\ul{n}_{\alpha_1},\ul{n}_{\alpha_2},
\cdots, \ul{n}_{\alpha_{|\Gamma'|}})$.
We will sometimes abbreviate $(\ndot)_{\Gamma}$ by $\ndot$ and 
$(\undot)_{\Gamma}$ by $\undot\;$. For each $j$, define $\Gamma_j$ to be
the set such that $(\ndot)_{\Gamma_j}$ is the occupation number vector
specifying the state of node $\ul{x}_j$. Thus,  $x_j=(\ndot)_{\Gamma_j}$ and
$\ul{x}_j=(\undot)_{\Gamma_j}$.  If $\ul{x}_j$ is an external node, call
all the components of $(\ndot)_{\Gamma_j}$ {\it external occupation
numbers}, and  if $\ul{x}_j$ is an internal node, call all the components
of $(\ndot)_{\Gamma_j}$ {\it internal occupation numbers}. Let 
$\gex$ be the set of all $\alpha$ such that $n_{\alpha}$ is an external
occupation number, and let 
$\gin$ be the set of all $\alpha$ such that $n_{\alpha}$ is an internal
occupation number. Clearly, $\gin$ and $\gex$ are disjoint and their
union is $\Gamma$. Note that 
$(x_{\bolddot})_{\zex}=(\ndot)_{\gex}$,
$(x_{\bolddot})_{\zin}=(\ndot)_{\gin}$,
and $(x_{\bolddot})_{Z_N}=(\ndot)_{\Gamma}$ (Analogous statements
 with $x$ and $n$ underlined to indicate random variables also hold).

Consider a classical probability distribution $P(\ndot)$. For any
$\Gamma_0 \subset \Gamma$, we may define the characteristic function
$\chi_c$ by

\beq
\chi_c[(\ndot)_{\Gamma_0}]=
\sum_{\mdot} P(\mdot)\prod_{\alpha\in \Gamma_0}
\delta(m_{\alpha},n_{\alpha})
\;,
\eqlabel{4.6}\eeq
where $\delta$ is the Kronecker delta function. (The ``c" subscript in
$\chi_c$ stands for ``classical".) If $\Gamma_H$ and $\Gamma_E$ are
disjoint subsets of $\Gamma$, one defines the classical conditional
probability that $(\undot)_{\Gamma_H}=(\ndot)_{\Gamma_H}$ (hypothesis)
given that $(\undot)_{\Gamma_E}=(\ndot')_{\Gamma_E}$ (evidence), by 

\beq
P\left[ (\undot)_{\Gamma_H}=(\ndot)_{\Gamma_H} |
(\undot)_{\Gamma_E}=(\ndot')_{\Gamma_E} \right]\;\;=\;\;
\frac{ \chi_c[(\ndot)_{\Gamma_H}, (\ndot')_{\Gamma_E}] }
{ \chi_c[ (\ndot')_{\Gamma_E} ] }
\;.
\eqlabel{4.7}\eeq
By setting $\Gamma_E=\phi$ in the last equation, we  conclude that for
any $\Gamma_0\subset \Gamma$, 

\beq
P\left[ (\undot)_{\Gamma_0}=(\ndot)_{\Gamma_0} \right]\;\;=\;\;
\chi_c[(\ndot)_{\Gamma_0} ]
\;.
\eqlabel{4.8}\eeq
According to the definition Eq.(4.6), if $\Gamma_0$ and $\Gamma_0'$ are
disjoint subsets of $\Gamma$, then

\beq
\sum_{(\ndot')_{\Gamma_0'}}
\chi_c[(\ndot)_{\Gamma_0}, (\ndot')_{\Gamma_0'} ]=
\chi_c[(\ndot)_{\Gamma_0} ]
\;.
\eqlabel{4.9}\eeq
Equations (4.8) and (4.9) imply that
 
\beq
\sum_{(\ndot')_{\Gamma_0'}}
P[(\ndot)_{\Gamma_0}, (\ndot')_{\Gamma_0'} ]=
P[(\ndot)_{\Gamma_0} ]
\;,
\eqlabel{4.10}\eeq
for $\Gamma_0,\Gamma_0'\subset \Gamma$,
$\Gamma_0\cap \Gamma_0'=\phi$. Equation (4.9), when applied to
Eq.(4.7), yields

\beq
\sum_{(\ndot)_{\Gamma_H}}
P[(\ndot)_{\Gamma_H}| (\ndot')_{\Gamma_E} ]=
1
\;.
\eqlabel{4.11}\eeq
Furthermore, from Eq.(4.7) it is obvious that 

\beq
P[(\ndot)_{\Gamma_H}| (\ndot')_{\Gamma_E} ]\geq 0
\;.
\eqlabel{4.12}\eeq
 Equations (4.11) and (4.12) imply that Eq.(4.7) is an adequate
definition of conditional probability.

Now consider a quantum mechanical probability amplitude $A(\ndot)$. For
any $\Gamma_0\subset\Gamma$, we define the characteristic function $\chi$
by

\beq
\chi[(\ndot)_{\Gamma_0}]=
\sum_{(\mdot)_{\gex}}
\left |
\sum_{(\mdot)_{\gin}}
A(\mdot)
\prod_{\alpha\in\Gamma_0}
\delta(m_{\alpha}, n_{\alpha})
\right|^2
\;.
\eqlabel{4.13}\eeq
If $\Gamma_H$ and $\Gamma_E$ are
disjoint subsets of $\Gamma$, one defines the quantum mechanical
conditional probability that $(\undot)_{\Gamma_H}=(\ndot)_{\Gamma_H}$
(hypothesis) given that $(\undot)_{\Gamma_E}=(\ndot')_{\Gamma_E}$
(evidence), by 

\beq
P\left[(\undot)_{\Gamma_H}=(\ndot)_{\Gamma_H} |
(\undot)_{\Gamma_E}=(\ndot')_{\Gamma_E}\right]\;\;=\;\;
\frac{ \chi[(\ndot)_{\Gamma_H}, (\ndot')_{\Gamma_E}] }
{ 
\sum_{ (\mdot)_{\Gamma_H} } 
\chi[(\mdot)_{\Gamma_H}, (\ndot')_{\Gamma_E}]
}
 \;.
\eqlabel{4.14}\eeq
Note that the denominator of the right hand side of Eq.(4.14) depends on
both $\Gamma_E$ and $\Gamma_H$, unlike the analogous denominator in the
classical definition Eq.(4.7). In quantum mechanics,
with $\chi_c$ replaced by $\chi$, 
Eq.(4.8) is not true for all
$\Gamma_0\subset\Gamma$ (although it is true if
$\Gamma_0\subset\gex$).  In quantum
mechanics,  with $\chi_c$ replaced by $\chi$, Eqs.(4.9) and (4.10) are not
necessarily true, whereas Eqs.(4.11) and (4.12) are obviously always
true. Because Eqs.(4.11) and (4.12) are satisfied in quantum mechanics,
Eq.(4.14) is an adequate definition of conditional probability.

Note that both the classical and quantum mechanical
definitions of conditional probability Eqs. (4.7) and (4.14) 
implicitly assume that we perform non-destructive measurements
on any internal nodes that might be measured. Hence, if a particle
is detected at an internal node, it is allowed to continue past that
node so that it can reach  other nodes further downstream. If one
wishes to perform a destructive measurement on a particle when it
passes through an internal node ${\ul x}$, then one is representing the
physical situation by the wrong CB or QB net; what is required is a net
that has the node {\ul x} as an external node. 

In Appendix A, the classical and quantum mechanical definitions of
conditional probability Eqs.(4.7) and (4.14) are generalized so that
they allow either sharp or fuzzy hypotheses and pieces of evidence. If
the evidence  narrows the set of possible values for $\ul{n}_1$ to a
single number,   say $\ul{n_1}=1$, then we  say that the evidence for
$\ul{n}_1$ is {\it sharp}. If the evidence doesn't do this, then
 we  say that the evidence for 
$\ul{n}_1$ is {\it fuzzy}. Sharp and fuzzy hypotheses are defined
analogously.

Note that if we had used Eq.(4.7) (with $\chi_c$ replaced by $\chi$) as
the quantum mechanical definition of conditional probability, there
would have been no guarantee that Eq.(4.11) would be satisfied. 
For given $\Gamma_H,\Gamma_E$ and $(\ndot')_{\Gamma_E}$, define the
{\it quantum non-additivity factor} $f_{qna}$ by

\beq
f_{qna}[\Gamma_H,\Gamma_E,(\ndot')_{\Gamma_E}]=
\frac{
\sum_{ (\mdot)_{\Gamma_H} } 
\chi[(\mdot)_{\Gamma_H}, (\ndot')_{\Gamma_E}] }
{\chi[(\ndot')_{\Gamma_E}] }
\;.
\eqlabel{4.15}\eeq
This quantity will be calculated for the examples in the next section. If
$f_{qna}=1$, then Eq.(4.7) (with $\chi_c$ replaced by $\chi$) 
and Eq.(4.14) agree; if $f_{qna}\neq 1$, then Eq.(4.7) does not give a
well defined probability distribution for $(\ndot)_{\Gamma_H}$
whereas Eq.(4.14) does.

In quantum mechanics, 
$\chi[(\ndot)_{\Gamma_H}, (\ndot')_{\Gamma_E}]$
is proportional to the number of occurrences of 
$(\undot)_{\Gamma_H}=(\ndot)_{\Gamma_H}$
and  $(\undot)_{\Gamma_E}=(\ndot')_{\Gamma_E}$ in an experiment that
measures all the $\Gamma_H$ and  $\Gamma_E$ nodes.  
$\chi[ (\ndot')_{\Gamma_E}]$
is proportional to the number of occurrences of 
$(\undot)_{\Gamma_E}=(\ndot')_{\Gamma_E}$ in an experiment that
measures only the $\Gamma_E$ nodes, leaving the $\Gamma_H$ nodes
undisturbed. Thus, Eq.(4.7) 
(with $\chi_c$ replaced by $\chi$) refers to the ratio of the number of
occurrences in two different types of experiments (one type measuring
the $\Gamma_H\cup\Gamma_E$ nodes and the other only the 
 $\Gamma_E$ nodes). On the other hand, Eq.(4.14) refers to the ratio of
occurrences in a single type of experiment (measuring 
the $\Gamma_H\cup\Gamma_E$ nodes). It seems to us that the latter ratio
is the more useful of the two.

By using CB and QB nets, one is led easily and naturally to 
calculate  probabilities by considering sums over paths, the type of sums
advocated by Feynman for quantum mechanics and by Kac for Brownian motion.
Indeed, one can express the classical and quantum mechanical definitions 
of conditional probability Eqs.(4.7) and (4.14) 
 in terms of sums over paths rather than sums over
node states. We do so for  arbitrary CB and QB nets in Appendix B.
Specific examples illustrating the affinity of QB nets with sums over
paths can be found in the next section and in Appendix C. Appendix C
presents a QB net which yields the Feynman path integral for a single
mass particle under the influence of an arbitrary potential. Using an
approach similar to that of Appendix C, it should be possible to define
QB nets that yield the Feynman integrals employed in non-relativistic and
relativistic quantum field theories.

\section*{5.  EXAMPLES OF QB NETS}

In this section, we will present the results of a computer program that
calculates conditional probabilities for QB nets. In particular, we
shall consider QB nets for experiments containing either two or three
Stern-Gerlach magnets. We will restrict our attention to experiments
involving a single particle. For simplicity, 
for each experimental configuration, we will assume 
 that the magnetic field vectors of all the
Stern-Gerlach magnets are coplanar.

Figures 15, 16 and 17 show the three kinds of nodes
that will be used in this section.

 In Fig.15, the triangle  represents a root node. This node will stand
for  $\psi_{n_{z-}n_{z+}}$, with 
$(n_{z-},n_{z+})\in\{ (0,1), (1,0)\}$ and
$|\psi_{01}|^2 +  |\psi_{10}|^2=1$.
 $\psi_{n_{z-}n_{z+}}$  is just the initial wavefunction for the single
particle under consideration. 

In Fig.16a, the black-filled circle  represents a  {\it marginalizer
node}.  The single incoming arrow is in a  state characterized by a vector
$(n_1,n_2,\cdots,n_K)$ of occupation numbers. The outgoing arrow is in a
state   characterized by a single occupation number $n_1'$. The
amplitude associated with the node is 

\beq
A(n_1' | n_1,n_2,\cdots,n_K)=\delta(n_1',n_1)
\;.
\eqlabel{5.1a}\eeq
 Thus, a  marginalizer node
takes a vector of occupation numbers and projects out one of its
components. Note that Eq.(5.1a) satisfies Eq.(4.1).

In Fig.16b, the black-filled circle  with a phase factor next to
it represents a  {\it phase shifter node}.  The single incoming 
(outgoing) arrow is
in a  state characterized by a single  occupation
number $n_1$ ($n'_1$).  The amplitude associated with the node
is 

\beq
A(n_1' | n_1)=e^{i\xi}\delta(n_1',n_1)
\;,
\eqlabel{5.1b}\eeq
where $\xi$ is a real constant.
 Note that Eq.(5.1b) satisfies Eq.(4.1).

In Figs.17, the  white-filled circles 
represent Stern-Gerlach magnets. The outgoing arrows are
labelled by a vector  ${\bf n}_u^{\alpha}=(n_{u-}^{\alpha},
n_{u+}^{\alpha})$ of occupation  numbers. Here the unit vector $\hat{u}$
represents the direction of the magnet's magnetic field, and $\alpha$
labels the magnet ( $\hat{u}$ is not enough to label the magnet if the
experiment contains more that one magnet whose magnetic field points
along the $\hat{u}$ direction.) The vector ${\bf n}_u^{\alpha}\in \{(0,0),
(0,1),(1,0)\}$ specifies a state (see Appendix D)

\beq
(a_{u-}^{\alpha \dagger})^{n_{u-}^{\alpha}}
(a_{u+}^{\alpha \dagger})^{n_{u+}^{\alpha}} 
\ket{0}
\;.
\eqlabel{5.2}\eeq
The creation operators $a_{u-}^{\alpha \dagger}$ and 
$a_{u+}^{\alpha \dagger}$ create particles in the $\ket{-_u}$ and
$\ket{+_u}$ states respectively. In  Fig.17a there is one
arrow entering the node, whereas in the Fig.17b there are two. In general,
since we are considering a single particle experiment, there may be any
number of arrows entering the node, but only one may be in state 1, all
others must be in state 0. In Figs.17a and 17b, the amplitude assigned to
each node is given by a table next to the graph. 
Clearly, the tables in
Figs.17a and 17b both satisfy Eq.(4.1).
For each experimental configuration,
we will assume  that the magnetic
fields of all magnets are coplanar. 
The plane containing these magnetic field
vectors may be chosen to be the X-Z plane with $\phi=0$.  
Thus, the matrix elements in the tables of Figs.17a and 17b 
are given in terms of angular parameters by Eqs.(D.7) and
(D.8).
  
(a) {\sc experiments with 2 stern-gerlach magnets} 

We will consider 2 configurations with 2 magnets: Fig.18 (the tree
diagram) and Fig.19 (the loop diagram).
 The diagrams are like road maps, with the arrows representing
the various roads along which the particle may travel.

A single particle travelling through the experimental configuration of
Fig.18 could exit through either the $\ul{n}_{z-}$, $\ul{n}_{u-}$ or
$\ul{n}_{u+}$ nodes. There is only one possible path leading to each of
these outcomes. Thus, one has

\beq
\begin{array}{l}
FI_1 = A(\pi_1)
\;,\\
\;\;\;\;A(\pi_1)=\psi_{10}
\;,\\
FI_2 = A(\pi_2)
\;,\\
\;\;\;\;A(\pi_2)=\av{-_u |+_z}\psi_{01}
\;,\\
FI_3 = A(\pi_3)
\;,\\
\;\;\;\;A(\pi_3)=\av{+_u |+_z}\psi_{01}
\;.
\end{array}
\eqlabel{5.3}\eeq
 For
$i\in\{1,2,3\}$, $A(\pi_i)$ is the amplitude for path $\pi_i$. For 
$j\in\{1,2,3\}$, $FI_j$ is a Feynman integral; that is, the sum of the
amplitudes for all paths with a given final state. (See Appendix B). We
have already checked that Eq.(4.1) is satisfied. We did so when we
defined the wavefunction, marginalizer and Stern-Gerlach nodes.  It is 
easy to show that $\sum_{j=1}^3 |FI_j|^2=1$, so Eq.(4.3) is also
satisfied for this net.    

A single particle travelling through the experimental configuration of
Fig.19 could exit through either the $\ul{n}_{u-}$ or the $\ul{n}_{u+}$ 
nodes. There are two possible paths leading to each of these two
outcomes. Thus, one has

\beq
\begin{array}{l}
FI_1 = A(\pi_1) + A(\pi_2)
\;,\\
\;\;\;\;A(\pi_1)=\av{-_u |-_z}\psi_{10}
\;,\\
\;\;\;\;A(\pi_2)=\av{-_u |+_z}\psi_{01}
\;,\\
FI_2 = A(\pi_3) + A(\pi_4)
\;,\\
\;\;\;\;A(\pi_3)=\av{+_u |-_z}\psi_{10}
\;,\\
\;\;\;\;A(\pi_4)=\av{+_u |+_z}\psi_{01}
\;.
\end{array}
\eqlabel{5.4}\eeq
As before, for
$i\in\{1,2,3\}$, $A(\pi_i)$ is the amplitude for path $\pi_i$, and  for 
$j\in\{1,2\}$, $FI_j$ is a Feynman integral. It is  easy to
show that $\sum_{j=1}^2 |FI_j|^2=1$, so Eq.(4.3) is satisfied by this
net. 

We will call an arrow or a node {\it simple} if its state is
characterized by a single occupation number (for example, a marginalizer
node and its outgoing arrow are both simple). We will say that a net is 
{\it fully marginalized} if  the incoming arrows of any node that is
not a marginalizer are all simple, and all external
arrows are simple. The nets of Figs.18 and 19, and, in fact, all the nets
considered in this section, are fully marginalized.

For each of the nets of Figs.18 and 19, we used a computer program to
calculate  probabilities with  one or two hypotheses
and with zero, one or two pieces of evidence.  More precisely, 
for the 2 QB nets in Figs. 18 and 19, and for their parent
CB nets, for  $\ul{n}$, $\ul{n}'$, $\ul{m}$ and 
$\ul{m}'\in\{ \ul{n}_{u+}, \ul{n}_{u-}, \ul{n}_{z+}, \ul{n}_{z-}\}$, 
for all $n$, $n'$, $m$ and $m'\in\{0,1\}$, we calculated (using
Eqs.(4.7) or (4.14)) the unconditional probabilities $P(n)$ and
$P(n,n')$, and the following conditional probabilities:

\beq
P(\ul{n}=n | \ul{m}=m )
\;,
\eqlabel{5.5}\eeq

\beq
P(\ul{n}=n | \ul{m}=m, \ul{m}'=m' )
\;,
\eqlabel{5.6}\eeq

\beq
P(\ul{n}=n, \ul{n}'=n' | \ul{m}=m )
\;,
\eqlabel{5.7}\eeq

\beq
P(\ul{n}=n, \ul{n}'=n' | \ul{m}=m, \ul{m}'=m' )
\;.
\eqlabel{5.8}\eeq

The table of Fig.20, call it the {\it evidence-case file}, gives the
various sets of evidence that were considered. For example, in case 2
(the row that starts with a 2), we assumed $n_{z+}=0$, whereas the
values of the remaining occupation numbers ( $n_{z-}, n_{u+},n_{u-} $)
were assumed to be unknown (i.e.,  we assumed there was no evidence as to
whether they were  0 or 1 and this is indicated in
Fig.20 with a blank space  ). In case 11, we assumed $n_{z-}=1$ and
$n_{z+}=0$, whereas the values of the remaining occupation numbers ( 
$n_{u+},n_{u-} $) were assumed to be unknown. Note that for the
evidence-cases 2 to 9 we considered 1 piece of evidence (as in
Eqs.(5.5) and (5.7)), and  for the evidence-cases 10 to 33 we
considered 2 pieces of evidence (as in Eqs.(5.6) and (5.8)).

To get numerical values 
for the probabilities associated with the nets
 Figs.18 and 19, particular values
had to be assumed for the initial wavefunction  and for the magnetic
field direction of each magnet. We used 

\beq
\psi_{01}=\frac{1+i}{2}
\;,\;\;
\psi_{10}=\frac{1}{\sqrt{2}}
\;,\;\;
\theta_u-\theta_z= \frac{\pi}{5}
\;.
\eqlabel{5.9}\eeq

For the tree graph of Fig.18, all the probability distributions for the
QB net and for its parent CB net were identical. This is not surprising
since {\it tree graphs}, i.e., graphs without any loops,
have no interfering paths. That is, for a given final state, there is
only one possible path that produces that final state.
More interesting  results were obtained for the loop graph of Fig.19.

Figures 21 and 22 show our computer program's  output 
 for the loop net,
for 
  cases 1, 2, 4, 10,  12 (See evidence-case
file  Fig.20).  Figure 21 gives one-hypothesis probabilities of the type
Eq.(5.5) and (5.6), whereas Fig.22 gives two-hypotheses probabilities of
the type Eq.(5.7) and (5.8).

First consider Fig.21. Columns A to D refer to the parent CB net and
columns F to I to the QB net.  Column A for the CB net (F for the QB
net) gives the identity of the occupation number $\ul{n}$ in Eqs.(5.5)
and (5.6). Columns B and C for the CB net (G and H for the QB
net) give the probabilities that $\ul{n}=0$ and $\ul{n}=1$, respectively,
in light of the evidence.  Column D for the CB net (I for the QB net)
gives the quantity $f_{qna}$ defined by Eq.(4.15). Note that for
evidence-case 10, there was no output, because the computer program
detected a contradiction. (In case 10, we were assuming that
$n_{z+}=n_{z-}=0$, which is impossible   since $n_{z+}+n_{z-}=1$.). More
interesting is the evidence-case 4. In this case, for the QB net, column
I indicates that $f_{qna}\neq 1$ for the distributions $P(n_{z+} |
\ul{n}_{u+}=0)$ and  $P(n_{z-} | \ul{n}_{u+}=0)$.\enote{pre-conditioning}

Next consider Fig.22. Columns A to F  refer to the
parent CB net and columns H to M to the QB net.  Column A for the CB net
(H for the QB net) gives the identity of the occupation numbers
$\ul{n}$ and $\ul{n}'$ in Eqs.(5.7) and (5.8). Columns B to E for the
CB net (I to L for the QB net) give the probabilities that 
$(\ul{n}, \ul{n}')=
(0,0), (0,1), (1,0), (1,1) $, respectively. Column F for the CB net (M
for the QB net) gives $f_{qna}$. For example, in Fig.22 we see that for
evidence-case 4, for the QB net,  all the probability distributions
except  $P( n_{u+}, n_{u-} |
\ul{n}_{u+}=0)$  
have $f_{qna}\neq 1$.\enote{pre-conditioning}

(b) {\sc experiments with three stern-gerlach magnets}

We will consider 7 configurations with 3 Stern-Gerlach magnets:
 Figs.23 to 29.

For Fig.23, define

\beq
\begin{array}{l}
FI_1 = A(\pi_1)
\;,\\
\;\;\;\;A(\pi_1)=\psi_{10}
\;,\\
FI_2 = A(\pi_2)
\;,\\
\;\;\;\;A(\pi_2)=\av{-_v |+_z}\psi_{01}
\;,\\
FI_3 = A(\pi_3)
\;,\\
\;\;\;\;A(\pi_3)=\av{-_u |+_v}\av{+_v |+_z}\psi_{01}
\;,\\
FI_4 = A(\pi_4)
\;,\\
\;\;\;\;A(\pi_4)=\av{+_u |+_v}\av{+_v |+_z}\psi_{01}
\;.
\end{array}
\eqlabel{5.10}\eeq
It is  easy to
show that $\sum_{j=1}^4 |FI_j|^2=1$, so Eq.(4.3) is satisfied by this
net.

For Fig.24, define

\beq
\begin{array}{l}
FI_1 = A(\pi_1) + A(\pi_2)
\;,\\
\;\;\;\;A(\pi_1)=\av{-_v |-_z}\psi_{10}
\;,\\
\;\;\;\;A(\pi_2)=\av{-_v |+_z}\psi_{01}
\;,\\
FI_2 = A(\pi_3) + A(\pi_4)
\;,\\
\;\;\;\;A(\pi_3)=\av{-_u |+_v}\av{+_v |-_z}\psi_{10}
\;,\\
\;\;\;\;A(\pi_4)=\av{-_u |+_v}\av{+_v |+_z}\psi_{01}
\;,\\
FI_3 = A(\pi_5) + A(\pi_6)
\;,\\
\;\;\;\;A(\pi_5)=\av{+_u |+_v}\av{+_v |-_z}\psi_{10}
\;,\\
\;\;\;\;A(\pi_6)=\av{+_u |+_v}\av{+_v |+_z}\psi_{01}
\;.
\end{array}
\eqlabel{5.11}\eeq
It is  easy to
show that $\sum_{j=1}^3 |FI_j|^2=1$, so Eq.(4.3) is satisfied by this
net. 

For Fig.25, define

\beq
\begin{array}{l}
FI_1 = A(\pi_1) 
\;,\\
\;\;\;\;A(\pi_1)=\psi_{10}
\;,\\
FI_2 = A(\pi_2) + A(\pi_3)
\;,\\
\;\;\;\;A(\pi_2)=\av{-_u |-_v}\av{-_v |+_z}\psi_{01}
\;,\\
\;\;\;\;A(\pi_3)=\av{-_u |+_v}\av{+_v |+_z}\psi_{01}
\;,\\
FI_3 = A(\pi_4) + A(\pi_5)
\;,\\
\;\;\;\;A(\pi_4)=\av{+_u |-_v}\av{-_v |+_z}\psi_{01}
\;,\\
\;\;\;\;A(\pi_5)=\av{+_u |+_v}\av{+_v |+_z}\psi_{01}
\;.
\end{array}
\eqlabel{5.12}\eeq
It is  easy to
show that $\sum_{j=1}^3 |FI_j|^2=1$, so Eq.(4.3) is satisfied by this
net.

For Fig.26, define

\beq
\begin{array}{l}
FI_1 = A(\pi_1) + A(\pi_2) + A(\pi_3) + A(\pi_4)  
\;,\\
\;\;\;\;A(\pi_1)=\av{-_u |+_v}\av{+_v |-_z}\psi_{10}
\;,\\
\;\;\;\;A(\pi_2)=\av{-_u |-_v}\av{-_v |-_z}\psi_{10}
\;,\\
\;\;\;\;A(\pi_3)=\av{-_u |+_v}\av{+_v |+_z}\psi_{01}
\;,\\
\;\;\;\;A(\pi_4)=\av{-_u |-_v}\av{-_v |+_z}\psi_{01}
\;,\\
FI_2 = A(\pi_5) + A(\pi_6) + A(\pi_7) + A(\pi_8)  
\;,\\
\;\;\;\;A(\pi_5)=\av{+_u |+_v}\av{+_v |-_z}\psi_{10}
\;,\\
\;\;\;\;A(\pi_6)=\av{+_u |-_v}\av{-_v |-_z}\psi_{10}
\;,\\
\;\;\;\;A(\pi_7)=\av{+_u |+_v}\av{+_v |+_z}\psi_{01}
\;,\\
\;\;\;\;A(\pi_8)=\av{+_u |-_v}\av{-_v |+_z}\psi_{01}
\;.
\end{array}
\eqlabel{5.13}\eeq
It is  easy to
show that $\sum_{j=1}^2 |FI_j|^2=1$, so Eq.(4.3) is satisfied by this
net. 

For Fig.27, define

\beq
\begin{array}{l}
FI_1 = A(\pi_1)
\;,\\
\;\;\;\;A(\pi_1)=\av{-_v |-_z}\psi_{10}
\;,\\
FI_2 = A(\pi_2)
\;,\\
\;\;\;\;A(\pi_2)=\av{+_v |-_z}\psi_{10}
\;,\\
FI_3 = A(\pi_3)
\;,\\
\;\;\;\;A(\pi_3)=\av{-_u |+_z}\psi_{01}
\;,\\
FI_4 = A(\pi_4)
\;,\\
\;\;\;\;A(\pi_4)=\av{+_u |+_z}\psi_{01}
\;.
\end{array}
\eqlabel{5.14}\eeq
It is  easy to
show that $\sum_{j=1}^4 |FI_j|^2=1$, so Eq.(4.3) is satisfied by this
net. 

For Fig.28, define

\beq
\begin{array}{l}
FI_1 = A(\pi_1) 
\;,\\
\;\;\;\;A(\pi_1)=\av{-_v |-_z}\psi_{10}
\;,\\
FI_2 = A(\pi_2) + A(\pi_3)
\;,\\
\;\;\;\;A(\pi_2)=e^{i \xi} \av{+_u |+_v}\av{+_v |-_z}\psi_{10}
\;,\\
\;\;\;\;A(\pi_3)=\av{+_u |+_z}\psi_{01}
\;,\\
FI_3 = A(\pi_4) + A(\pi_5)
\;,\\
\;\;\;\;A(\pi_4)=e^{i \xi} \av{-_u |+_v}\av{+_v |-_z}\psi_{10}
\;,\\
\;\;\;\;A(\pi_5)=\av{-_u |+_z}\psi_{01}
\;.
\end{array}
\eqlabel{5.15}\eeq
Note that we have included a phase factor $e^{i \xi}$ which can be
thought of as arising from the Stern-Gerlach node $u$ when its input
$n_{v+}$ equals 1 (or from node $v$ when
its output $n_{v+}$ equals 1, or from node $z$ when its output $n_{z-}$ 
equals
1). Alternatively, the phase factor may be though of as arising from a
phase shifter node, not pictured in Fig.28, located in the middle of the
arrow labelled $n_{v+}$.  If $\xi=0$, then  $\sum_{j=1}^3 |FI_j|^2 \neq
1$ and therefore Eq.(4.3)  is violated. On the other hand, if

\beq
e^{i\xi}= i \frac{ \psi_{01}\psi_{10}^* }
{|\psi_{01}\psi_{10}^*|}
\;,
\eqlabel{5.16}\eeq
then $\sum_{j=1}^3 |FI_j|^2=1$.\enote{pf source}

For Fig.29, define

\beq
\begin{array}{l}
FI_1 = A(\pi_1) + A(\pi_2) + A(\pi_3)  
\;,\\
\;\;\;\;A(\pi_1)=\av{+_u |+_z}\psi_{01}
\;,\\
\;\;\;\;A(\pi_2)=e^{i \xi} \av{+_u |+_v}\av{+_v |-_z}\psi_{10}
\;,\\
\;\;\;\;A(\pi_3)=e^{i \xi} \av{+_u |-_v}\av{-_v |-_z}\psi_{10}
\;,\\
FI_2 = A(\pi_4) + A(\pi_5) + A(\pi_6)
\;,\\
\;\;\;\;A(\pi_4)=\av{-_u |+_z}\psi_{01}
\;,\\
\;\;\;\;A(\pi_5)=e^{i \xi} \av{-_u |+_v}\av{+_v |-_z}\psi_{10}
\;,\\
\;\;\;\;A(\pi_6)=e^{i \xi} \av{-_u |-_v}\av{-_v |-_z}\psi_{10}
\;.
\end{array}
\eqlabel{5.17}\eeq
It is  easy to
show that $\sum_{j=1}^2 |FI_j|^2=1$ for any value of $\xi$. 
However, we believe the correct   choice of $\xi$ to be the one
given  by Eq.(5.16). For this choice, $P(\ul{n}_{u-}=1|
\ul{n}_{v-}=0)$ is the same for Figs.28 and 29. And this is what  we
expect. Otherwise, empty De Broglie waves could influence the outcome
of an experiment (and thus could be detected), which does not appear to
be the case experimentally.

For the  QB nets of Figs.23 to 29, and for their parent
CB nets, for  $\ul{n}$, $\ul{n}'$, $\ul{m}$ and 
$\ul{m}'\in\{ \ul{n}_{u\pm}, \ul{n}_{v\pm}, \ul{n}_{z\pm}\}$,  for all
$n$, $n'$, $m$ and $m'\in\{0,1\}$, we calculated the conditional
probabilities of Eqs.(5.5) to (5.8). 

The table of Fig.30  is an evidence-case file that gives the sets of
evidence that were considered for the nets of Figs.23 to 29. For example,
in case 2 ( the row that starts with a 2),
we assumed $n_{z+}=0$, whereas the values of the remaining 
occupation numbers ($n_{u\pm}, n_{v\pm}, n_{z-}$) were assumed to be
unknown.

To get numerical values for the probabilities associated with the nets
of Figs.23 to 29, particular values for 
$\psi_{01}, \psi_{10}, \theta_u, \theta_v, \theta_z$ were assumed.

For  the tree graphs Figs.23 and 27, and also for the non-tree graph 
Fig.28, we found that the QB net and its parent CB net yielded
identical probability distributions.

\section*{APPENDIX A. CONDITIONAL PROBABILITIES \\
                      FOR FUZZY MEASUREMENTS}

In this appendix,
we will generalize the classical  and quantum mechanical definitions
 of conditional probability
Eqs.(4.7) and (4.14) so as to include either
sharp or fuzzy hypotheses and pieces of evidence.

The following simple results from set theory are relevant.
Given any finite set $S$, we will denote the number of elements in $S$ 
 by $|S|$. Given two sets $R$ and $S$, we define the {\it direct
product set } $R\times S$  by

\beq
R\times S=
 \{(x,y)| x\in R, y\in S\}
\;.
\eqlabel{A.1}\eeq
$R\times S$ is also
denoted by the {\it vector of sets} $(R,S)$.  Given sets
$R_1,R_2,S_1,S_2$, it is easy to show that (see Fig.A.1)

\beq
(R_1\times R_2)\cap (S_1\times S_2)=
(R_1\cap S_1)\times (R_2\cap S_2)
\;.
\eqlabel{A.2}\eeq
Therefore, $(R_1\times R_2)\cap (S_1\times S_2)=\phi$ if and only  if
$R_1\cap S_1=\phi$ or $R_2\cap S_2=\phi$. Given any set $S$ and any
integer $n$, we will denote by $S^n$ the product
 $S\times S\times \cdots \times S$
of $n$ copies of $S$. 

Suppose $Q_{\alpha}\subset
\zop$ for each $\alpha$. If 
$\Gamma'=\{\alpha_1,\alpha_2,\cdots, \alpha_{|\Gamma'|}\}\subset \Gamma$,
we define the  vector of sets or direct product set 
$(Q_{\bolddot})_{\Gamma'}$  by $(Q_{\bolddot})_{\Gamma'}=
(Q_{\alpha_1},Q_{\alpha_2},\cdots,Q_{\alpha_{|\Gamma'|}} )$.
Equivalently, one may write 
$(Q_{\bolddot})_{\Gamma'}= 
Q_{\alpha_1}\times Q_{\alpha_2}\times
\cdots\times Q_{\alpha_{|\Gamma'|}}$. 
Note that 
$(Q_{\bolddot})_{\Gamma'} \subset \zop^{|\Gamma'|}$.
Sometimes,
we will abbreviate $(Q_{\bolddot})_{\Gamma}$ by just $Q_{\bolddot}$. 
Two direct product sets $(R_{\bolddot})_{|\Gamma'|}$ and 
$(S_{\bolddot})_{|\Gamma'|}$ are disjoint if and only if there exists an
$\alpha\in \Gamma'$ such that $R_{\alpha}\cap S_{\alpha}=\phi$.
We will need to consider collections 
$\calh =\{ \hdot^i | i=1,2\cdots,|\calh | \}$ of direct product
sets 
$\hdot^i \subset \zop^{|\Gamma|}$. Such a collection will
be said to be a {\it partition} of $\zop^{|\Gamma|}$ if  any pair of
distinct sets $\hdot^i$ and $H_{\bolddot}^j$ is disjoint and 
$\cup_{i=1}^{|\calh |} \hdot^i = \zop^{|\Gamma|}$. 

 We are interested in  $P(\undot\in \hdot|\undot\in\edot)$,
i.e., the probability that $\ul{n}_{\alpha}\in H_{\alpha}$ for all 
$\alpha\in \Gamma$ given that $\ul{n}_{\beta}\in E_{\beta}$ for all 
$\beta\in \Gamma$. $H_{\alpha}\subset \zop$ is the {\it hypothesis for }
$\ul{n}_{\alpha}$, and   $E_{\beta}\subset \zop$ is the {\it evidence
for} $\ul{n}_{\beta}$. If $H_{\alpha}$ contains only one element
of $\zop$, then we say that the hypothesis for $\ul{n}_{\alpha}$ is 
{\it sharp}, whereas if $H_{\alpha}$ contains more that one element of
$\zop$, we say that the hypothesis for $\ul{n}_{\alpha}$ is {\it fuzzy}.
Analogously, the number of elements in $E_{\beta}$ determines whether
the evidence for $\ul{n}_{\beta}$ is sharp or fuzzy.

For any set $S$, define the {\it indicator function}
$1_{S}(x)$ by 

\beq
1_S(x)=\left\{
\begin{array}{ll}
0& \mbox{ if $x\not\in S$}\\
1& \mbox{ if $x\in S$}
\end{array}
\right.
\;.
\eqlabel{A.3}\eeq

Consider first a classical probability distribution $P(\ndot)$. One
defines 

\beq
P(\undot\in \hdot|\undot\in\edot)=
\frac{ \sum_{\ndot\in \hdot\cap\edot} P(\ndot) }
{ \sum_{\ndot\in \edot} P(\ndot) }=
\frac{ \sum_{\ndot} P(\ndot) 
\prod_{\alpha\in\Gamma} 1_{H_{\alpha}\cap E_{\alpha}} (\nal) } 
{ \sum_{\ndot} P(\ndot) 
\prod_{\alpha\in\Gamma} 1_{E_{\alpha}} (\nal)}
\;.
\eqlabel{A.4}\eeq
To write the last equation more succinctly, it is convenient to define the
{\it filter function } $f_{Q_{\bolddot}}(\ndot)$, for any 
direct product set $Q_{\bolddot}\subset \zop^{|\Gamma|}$, by

\beq
 f_{Q_{\bolddot} }(\ndot)=
\prod_{\alpha\in\Gamma} 1_{Q_{\alpha}} (\nal)
\;.
\eqlabel{A.5}\eeq 
It is also convenient to define the {\it characteristic functional} 
$\chi_c[K]$ of any function $K(\ndot)$ of $\ndot$ by

\beq
\chi_c [K]= \sum_{\ndot} P(\ndot)K(\ndot)
\;.
\eqlabel{A.6}\eeq
(The $c$ subscript in $\chi  _c$ stands for ``classical"). Now Eq.(A.4)
can be written succinctly as 

\beq
P(\undot\in \hdot|\undot\in\edot)=
\frac{\chi_c [f_{\hdot\cap\edot}]}
{\chi_c[f_{\edot}]}
\;.
\eqlabel{A.7}\eeq
Note that $f_{\hdot\cap\edot}=f_{\hdot}f_{\edot}$.
Note that if $\calh=\{\hdot^i | i=1,2,\cdots,|\calh|\}$ is a partition
of $\zop^{|\Gamma|}$, then Eq.(A.7) satisfies

\beq
\sum_{i=1}^{|\calh|}
P(\undot\in \hdot^i|\undot\in\edot)=1
\;.
\eqlabel{A.8}\eeq

Now consider a  quantum mechanical probability amplitude $A(\ndot)$.
 We define the {\it characteristic
functional} $\chi  [K]$ by

\beq
\chi [K]= \sum_{(\ndot)_{\gex}}
\left |
\sum_{(\ndot)_{\gin}} 
A(\ndot)K(\ndot)
\right| ^2
\;.
\eqlabel{A.9}\eeq
Suppose  $\calh=\{\hdot^i | i=1,2,\cdots,|\calh|\}$ is a partition
of $\zop^{|\Gamma|}$. In quantum mechanics, the conditional
probability that $\undot\in \hdot^i$ given that $\undot\in \edot$,
depends on the choice of partition $\calh$. We define this probability
by

\beq
P_{\calh}(\undot\in \hdot^i |\undot\in\edot)=
\frac{\chi [f_{\hdot^i\cap\edot}]}
{\sum_{j=1}^{|\calh|} \chi [f_{\hdot^j\cap\edot}]}
\;.
\eqlabel{A.10}\eeq
Equation (A.8) is trivially satisfied by  Eq.(A.10).

\section*{APPENDIX B.  CONDITIONAL PROBABILITIES\\ EXPRESSED
                       IN TERMS OF PATH SUMS}

In this appendix, we will express the classical and quantum
mechanical definitions Eqs.(A.7) and (A.10) of conditional probability
in terms of sums over paths. Simple examples of the results of this
appendix may be found in Section 5 of this paper.

Let $P(\ndot)$ and $A(\ndot)$ be the values of a CB net
and a QB net,  respectively. Suppose that ${\cal P}(\ndot)=P(\ndot)$ if a
CB net is being considered and ${\cal P}(\ndot)=A(\ndot)$ if a QB net is.

Let $\Pi$ be  the set of all possible {\it paths}. $\Pi$ is defined so
that there is exactly one $\pi\in\Pi$ for each $\ndot$ that satisfies
${\cal P}(\ndot)\neq 0$. Thus, 
there is a one to one onto map $\ndot(\pi)$ from $\Pi$ into
$\{\ndot\;|{\cal P}(\ndot)\neq 0\}$.
 
For any $\Gamma'\subset \Gamma$, if $\mdot$ is such that 
$(\mdot)_{\Gamma'}=(\ndot)_{\Gamma'}$, 
call $\mdot$ an {\it extension} of 
$(\ndot)_{\Gamma'}$.
Let $\Sigma$ be 
the set of all possible {\it final states}. 
$\Sigma$ is 
defined so that there is
exactly one $\sigma\in\Sigma$ for each $(\ndot)_{\gex}$ for which
there exists an extension $\ndot$ such that  
${\cal P}(\ndot)\neq 0$. Thus, 
there is a one to one onto map 
from $\Sigma$ into $\{(\ndot)_{\gex}|{\cal P}(\ndot)\neq 0\}$. 

Define the {\it partition function} $Z$ to be a
function from $\Pi$ to $\Sigma$ such that $Z(\pi)=\sigma$ if the path
$\pi$ has $\sigma $ as final state. For each $\sigma\in \Sigma$, let
$C(\sigma)$ be the set  of all paths $\pi$ that have $\sigma$ as
final state; i.e., $C(\sigma) =\{ \pi\in \Pi | Z(\pi)=\sigma \}$.
Clearly, the sets (equivalence classes) $C(\sigma)$ and $C(\sigma')$ are
disjoint when $\sigma\neq\sigma'$, and
$\cup_{\sigma\in \Sigma} C(\sigma)=\Pi$.

For each $\pi\in \Pi$, define $\ol{{\cal P}}(\pi)$ by
$\ol{{\cal P}}(\pi)={\cal P}( \ndot(\pi) )$.

For any direct product set $Q_{\bolddot}\subset \zop^{|\Gamma|}$,
define the filter function $\ol{f}_{ Q_{\bolddot} }(\pi)$ by 

\beq
\ol{f}_{ Q_{\bolddot} }(\pi)=
\prod_{\alpha\in\Gamma} 1_{Q_{\alpha}}(n_{\alpha}(\pi))
\;.
\eqlabel{B.1}\eeq

Classically, one defines the characteristic functional $\ol{\chi}_c[K]$
of any function $K(\pi)$ of $\pi$ by

\beq
\ol{\chi}_c[K]=
\sum_{\sigma\in\Sigma}\;
\sum_{\pi\in C(\sigma)}
\ol{P}(\pi)K(\pi)
\;.
\eqlabel{B.2}\eeq
Then

\beq
P(\undot\in \hdot | \undot\in\edot)=
\frac{ \ol{\chi}_c(\ol{f}_{\hdot\cap\edot}) }
{ \ol{\chi}_c(\ol{f}_{\edot}) }
\;.
\eqlabel{B.3}\eeq

Quantum mechanically, one defines the characteristic functional
$\ol{\chi}[K]$ by

\beq
\ol{\chi}[K]=
\sum_{\sigma\in\Sigma}
\left|
\sum_{\pi\in C(\sigma)}
\ol{A}(\pi)K(\pi)
\right|^2
\;.
\eqlabel{B.4}\eeq
If $\calh=\{\hdot^i | i=1,2,\cdots,|\calh|\}$  is a partition 
 of $\zop^{|\Gamma|}$, then one defines

\beq
P_{\cal H}(\undot\in \hdot^i | \undot\in\edot)=
\frac{ \ol{\chi}(\ol{f}_{\hdot^i\cap\edot}) }
{ \sum_{j=1}^{|\calh | } \ol{\chi}(\ol{f}_{\hdot^j\cap\edot}) }
\;.
\eqlabel{B.5}\eeq

\section*{APPENDIX C.  QB NET FOR\\ 
SINGLE PARTICLE\\
 FEYNMAN PATH INTEGRAL}

In this appendix, we will present a QB net which yields the Feynman path
integral\enote{FH} that in turn yields 
the Schroedinger equation for a single mass particle in an arbitrary
potential. 

We begin by restricting positions $x$ to lie in the interval (``box") 
$[\frac{-L}{2}, \frac{L}{2}]$, where $L$ is  much larger that any
other length 
(including the particle's position coordinate) occurring in the problem.
 Then we divide the box into
subintervals of infinitesimal length $\Delta x$, and call the
midpoints of these subintervals $x_s$ with $s\in\{0, \pm 1, \pm
2,\cdots,\pm N_x/2\}=Z_x$.  Similarly,  
we restrict times $t$ to lie in the
interval $[0, T]$. Then we divide the interval $[0, T]$ into 
subintervals of infinitesimal length $\Delta t$, and call the midpoints
of these subintervals $t_i$ with $i\in\{0, 1,  2,\cdots,N_t\}=Z_t$.

 Henceforth, for any function $f(x_s)$,
we will use $f(\bolddot)$ to represent the vector whose $s$th
component is $f(x_s)$. For example,  $\delta_{x_r}(\bolddot)$ will
represent the vector whose $s$th component 
$\delta_{x_r}(x_s )$ is 1 if $r=s$ and zero otherwise. 

As in Fig.C.1, we assign a node and a 
random variable $\ul{n}(x_s,t_i)$ to all space-time lattice points
$(x_s,t_i)$ with $s\in Z_x$ and $i\in Z_t-\{0\}$ and to the lattice point
$(x_0,t_0)$. (Thus, at time $t_0=0$, only position $x_0$ gets a node).
We draw arrows pointing from each node to all nodes occurring a time 
$\Delta t$ later. 
We also draw external arrows pointing out of each node at time $t_{N_t}$.
The net starts at time $t_0$ with a single root node at position $x_0$,
and it ends at time $t_{N_t}$ with  external nodes at  
each position $x_s$ for all  $s\in Z_x$.
 The random variables 
$\ul{n}(x_s,t_i)$ are  occupation numbers that assume values 
$n(x_s,t_i)\in \zop$. These occupation numbers specify
 states 

\beq
\ket{n(x_s,t_i)}= 
\frac{[a^{\dagger }(x_s,t_i)]^{n(x_s,t_i)} } 	
{\sqrt{n(x_s,t_i)!} } \ket{0}
\;.
\eqlabel{C.1}\eeq
From the  states of Eq. (C.1), one can form states
$\ket{n(\bolddot,t_i)}$  defined
by

\beq
\ket{n(\bolddot,t_i)}= \Pi_{s\in Z_x} \ket{n(x_s,t_i)}
\;.
\eqlabel{C.2}\eeq

Figure C.2 shows the input and output arrows for a single node
$(x_s,t_i)$ of our net. This node is assigned a value

\beq
A[n(x_s,t_i)| n(\bolddot,t_{i-1})]=
\bra{n(x_s,t_i)} \Omega \ket{ n(\bolddot,t_{i-1})}
\;,
\eqlabel{C.3}\eeq
where $\Omega$ is a second quantized operator.

In general, the theory characterized by this QB net is a multi-particle
quantum field theory. In fact, $n(x_s,t_i)$ is the classical field
 $\phi(x_s,t_i)$ that is commonly used to define Feynman integrals in
quantum field theories. In this appendix, we will consider
a single particle. Thus, for all $i$,
  $\sum_{s\in Z_x} n(x_s,t_i)
=1$  and 
$n(\cdot,t_i)=\delta_{x_r}(\cdot)$ for some
$r\in Z_x$. For a single particle, the root node $(x_0,t_0)$ will be taken
to have amplitude 1 for $n(x_0,t_0)=1$ and amplitude 0 for
$n(x_0,t_0)=0$. All other nodes will be taken to have  the value 
$A[n(x_s,t_i)|  n({\bf  \cdot},t_{i-1})]$ that is  specified by the table
of Fig.C.3.  The quantity $\alpha_{s,r}$ in Fig.C.3  is given by

\beq
\alpha_{s,r}=
\bra{x_s} 
e^{\frac{-i}{\hbar} \Delta t\; H}
\ket{x_r}
\;.
\eqlabel{C.4}\eeq
In this last equation,
$\hbar$ is Planck's constant; 
$H$ is the single particle, first quantized Hamiltonian; 
$\ket{x_r}$ for $r\in Z_x$ are the position
eigenvectors, normalized so that
$\av{x_s|x_r}=\delta_{s,r}$
and $\sum_{r\in Z_x} \ket{x_r}\bra{x_r}=1 $,
where $\delta_{s,r}$ is the Kronecker delta. Equation (4.1)
is trivially satisfied by the table of Fig.C.3. As for
Eq.(4.3), it follows from the identity

\beq
\sum_{s\in Z_x} |\bra{x_ s}
e^{ \frac{-i}{\hbar}(t_{N_t}-t_0)H}
\ket{x_0}|^2 =1
\;.
\eqlabel{C.5}\eeq

For $H=\frac{p^2}{2m} + V(x,t)$, one may write

\beq
\alpha_{s,r}=
\exp \left\{
\frac{-i\Delta t}{\hbar}
\left [ (\frac{-\hbar^2}{2m})(\frac{d}{dx_s})^2 +
V(x_s,t) \right] \right\}
\Delta x \delta (x_s-x_r)
\;,
\eqlabel{C.6}\eeq
where $\delta(\cdot)$ is the Dirac delta function. If one were to expand
the exponential in the last equation, terms in which
$(\frac{d}{dx_s})^2$
 acted only on the delta function would be much larger than those in which
it acted on $V(x_s,t)$. Thus, one may approximate $\alpha_{s,r}$ by
keeping only those terms in which $(\frac{d}{dx_s})^2$
does not act on $V(x_s,t)$. One then gets

\beq
\alpha_{s,r}\approx (\alpha_{s,r})_{free}
\;e^{
\frac{-i}{\hbar}\Delta t V(x_s,t)}
\;,
\eqlabel{C.7}\eeq
where

\beq
(\alpha_{s,r})_{free}=\Delta x \int_{-\infty}^{+\infty} \frac{dk}{2\pi}
e^{ik(x_s-x_r) - \frac{i \Delta t}{\hbar} (\frac{\hbar^2k^2}{2m}) }
\;.
\eqlabel{C.8}\eeq
The Gaussian integral in Eq.(C.8) is easily performed. One
obtains

\beq
\alpha_{s,r}=\sqrt{\frac{-i \Delta \theta}{\pi}}
e^{\frac{i}{\hbar} \Delta t {\cal L}_{s,r} }
\;,
\eqlabel {C.9}\eeq
where

\beq
\Delta \theta =(\frac{\Delta t}{\hbar}) \frac{m}{2} 
(\frac{\Delta x}{\Delta t} )^2
\;,
\eqlabel{C.10}\eeq
and\enote{V} 

\beq
{\cal L}_{s,r} =
\frac{m}{2} 
\left(\frac{x_s-x_r}{\Delta t} \right)^2 - V(x_s,t)
\;.
\eqlabel{C.11}\eeq

In general, 
$|\alpha_{s,r}|<<1$. This is why. 
The phase $\frac{\Delta t {\cal L}_{s,r}}{\hbar}$ of $\alpha_{s,r}$ must not
show any granularity when $x_s$ and $x_r$ range over 
their possible discrete
values. Otherwise, the outcome of experiments  that depended on
this phase would depend on $\Delta x$.
When $x_s$
goes from $x_r$ to $x_r + \Delta x$,
the phase $\frac{\Delta t {\cal L}_{s,r}}{\hbar}$ jumps by $\Delta
\theta$. If this jump is small, then so is 
$|\alpha_{s,r}|=\sqrt{\frac{\Delta \theta}{\pi}}$.

 Consider Fig.C.4, which presents a typical evaluation of the net of
Fig.C.1. We have blackened those nodes whose state $\ul{n}(x_s,t_i)$ is
1. Arrows coming out of the blackened nodes are also imagined to lie
in state 1.  We show only those arrows lying in state 1; all other
arrows, i.e., those lying in state 0, are not shown. The value of each
node (the infinitesimal complex amplitude
$\alpha_{s,r}$ for the blackened nodes and approximately 1 for the
unblackened ones)
has been placed next to it. If one connects the blackened nodes one
obtains a path.  The value of the whole net equals the product of the node
values along this path. Thus, the sum over all possible evaluations of
Fig.C.1
 with  the same  final node (at time $T$) blackened 
 may be viewed as a 
sum over all paths with the same final state. This latter path sum
is precisely  the familiar Feynman path integral that yields the
Shroedinger equation for a single particle in an arbitrary
potential.\enote{FH}

Before concluding this appendix, let us consider the classical limit of
the QB net of Fig.C.1.

Recall the method of obtaining the particle's classical equation of
motion using the action. The action
$S[x(\cdot)]$ for a path $x(t)$ is

\beq
S[x(\cdot)]=\int_0^T dt \;{\cal L} 
\;,
\eqlabel{C.12}\eeq
where the Lagrangian ${\cal L}$ is

\beq
{\cal L}=\frac{m}{2} [\dot{x}(t)]^2 - V[x(t),t]
\;. 
\eqlabel{ C.13}\eeq
Let $\delta x(t)$ be any smooth function of
time. For any functional $F[x(\cdot)]$ of a function $x(t)$, define
$\delta F$ by

\beq
\delta F = F[x(\cdot)+\delta x(\cdot)]- F[x(\cdot)] 
\;.
\eqlabel{C.14}\eeq
The equation of motion is obtained by setting $\delta S=0$ to
first order in $\delta x(t)$ for all  $\delta x(\cdot)$ with $\delta
x(0)=\delta x(T)=0$. One gets

\beq
\delta S=\int_0^T  dt\;\left (
-m \ddot{x}(t) 
-\pder{V}{x}[x(t),t]\right) \delta x(t)
=0
\;,
\eqlabel{C.15}\eeq
for arbitrary $\delta x(t)$, which implies 

\beq
m \ddot{x}= 
-\pder{V}{x}
\;.
\eqlabel{C.16}\eeq
The classical path $x_{cl}(t)$ is defined as the solution to Eq.(C.16)
for given $x(0)$ and $x(T)$.  The classical action is defined by
$S_{cl}=S[x_{cl}(\cdot)]$. When $\delta S =0$, 
$x(t)=x_{cl}(t)$ and $S[x(\cdot)]=S_{cl}$.

If we restrict $x(t)$ to take on values in the set $\{x_r| r \in Z_x\}$,
then to the path $x(\cdot)$ there corresponds a sequence of integers
$r(t_i)$ for each $i\in Z_t$, where $x(t_i)=x_{r(t_i)}$. The path $x(t)$
connects a string of nodes $(x_{r(t_i)}, t_i)$ (the blackened nodes in
Fig.C.4). The value of the QB net Fig.C.1 is the the product of the values
$\alpha_{r(t_i),r(t_{i-1})}$ of these nodes.  From Eqs.(C.11), (C.12) and
(C.13), 

\beq
S[x(\cdot)]=\sum_{i\in Z_t} \Delta t {\cal L}_{r(t_i),r(t_{i-1})}
\;.
\eqlabel{C.17}\eeq
 Thus,

\beq
\prod_{i\in Z_t}
\alpha_{r(t_i),r(t_{i-1})} = 
\left( \frac{-i\Delta \theta}{\pi} \right)^{ \frac{N_t}{2} }
e^{ \frac{i}{\hbar} S[x(\cdot)] }
\;.
\eqlabel{C.18}\eeq
Probabilities are calculated by summing the weights Eq.(C.18)
for all possible paths $x(\cdot)$ with the same $x(T)$ and taking the
magnitude squared of the sum. The paths
which interfere constructively and therefore contribute the most to such
probabilities  are those for which 
$\left|\frac{\delta S}{\hbar}\right|<<1$,
 i.e.,  those for which the phase change $\frac{S}{\hbar}$
accumulated over the time interval $(0,T)$ is nearly stationary 
under deformations $\delta x(t)$ of the path.
In  the classical limit, which occurs 
when $\hbar\rarrow 0$, $\left|\frac{\delta S}{\hbar}\right|<<1$
 means $\delta S\rarrow 0$.
 
Note that the classical limit of the QB net Fig.C.1 is
not its parent CB net. With the parent CB net, paths do not interfere,
whereas with the QB net in the classical limit, there is so much
destructive interference between paths that all path except the classical
one cancel each other out.
With the parent CB
net, equal weight is given to smooth paths like those a classical
particle would follow and to jagged paths with unbounded
variations. With the QB net in the classical limit, paths with unbounded
variations cancel each other out.

\section*{APPENDIX D.  SPIN $\frac{1}{2}$ STATES}

Our conventions for spin $\frac{1}{2}$ states are as follows.

Let $\hat{u}$ be a unit 3-dimensional vector characterized by
angles  $(\theta, \phi)$ (see Fig.D.1). Thus,

\beq
\hat{u}=
\left[
\begin{array}{c}
\sin\theta \cos\phi\\
\sin\theta \sin\phi\\
\cos\theta
\end{array}
\right]
\;.
\eqlabel{D.1}\eeq
Let ${\bf \sigma}=(\sigma_x,\sigma_y,\sigma_z)$ be the vector of Pauli
matrices, where 

\beq
\sigma_x=
\left(
\begin{array}{cc}
0&1\\
1&0
\end{array}
\right)
\;,\;\;\;
\sigma_y=
\left(
\begin{array}{cc}
0&-i\\
i&0
\end{array}
\right)
\;,\;\;\;
\sigma_z=
\left(
\begin{array}{cc}
1&0\\
0&-1
\end{array}
\right)
\;.
\eqlabel{D.2}\eeq
If $\ket{+_u}$ and $\ket{-_u}$ are defined by

\beq
{\bf \sigma}{\cdot} \hat{u} \ket{+_u}=\ket{+_u}
\;,\;\;
{\bf \sigma}{\cdot} \hat{u} \ket{-_u}=-\ket{-_u}
\;,
\eqlabel{D.3}\eeq
then one can show that

\beq
\ket{+_u}=
\left(
\begin{array}{c}
C E^*\\
S E
\end{array}
\right)
\;,\;\;\;
\ket{-_u}=
\left(
\begin{array}{c}
-S E^*\\
C E
\end{array}
\right)
\;,
\eqlabel{D.4}\eeq
where 

\beq
S=\sin\frac{\theta}{2}
\;,\;\;
C=\cos\frac{\theta}{2}
\;,\;\;
E=e^{\frac{i\phi}{2} }
\;.
\eqlabel{D.5}\eeq
For example, if $\theta=\phi=0$, one gets

\beq
\ket{+_z}=
\left(
\begin{array}{c}
1\\
0
\end{array}
\right)
\;,\;\;\;
\ket{-_z}=
\left(
\begin{array}{c}
0\\
1
\end{array}
\right)
\;.
\eqlabel{D.6}\eeq

By Eqs.(D.4) and (D.5), if $\hat{u}$ and $\hat{u'}$ are unit vectors
characterized by angles $(\theta_u,\phi_u)$ and
$(\theta_{u'},\phi_{u'})$, respectively, and if $\phi_u=\phi_{u'}=0$, then

\beq
\av{+_{u'} | +_u}=
\av{-_{u'} | -_u}=
\cos\left( \frac{\theta_{u'}-\theta_u}{2} \right)
\;,
\eqlabel{D.7}\eeq

\beq
\av{+_{u'} | -_u}=
-\av{-_{u'} | +_u}=
\sin\left( \frac{\theta_{u'}-\theta_u}{2} \right)
\;.
\eqlabel{D.8}\eeq

For any unit vector $\hat{u}$, the antisymmetric state for two particles 1
and 2
is

\beq
\ket{\psi^{ant}_u}=
\frac{1}{\sqrt{2} }
\left[
\ket{+_u}_1\ket{-_u}_2-
\ket{-_u}_1\ket{+_u}_2
\right]
\;.
\eqlabel{D.9}\eeq
($\ket{\psi^{ant}_u}$ has zero total angular momentum and is thus also
called the singlet state). By expressing the states 
$\ket{\pm_u}$ on the right hand side of Eq.(D.9) in terms of  
$\ket{\pm_z}$, it is easy to show that
$\ket{\psi^{ant}_u}$ is invariant under rotations, i.e., it is independent
of $\hat{u}$.

So far, our formulation has been a first quantized one. In a second
quantized formulation, one defines annihilation operators
$a_{u \sigma}$,  for any unit vector $\hat{u}$ 
and for $\sigma\in\{+,-\}$ .
These annihilation operators must satisfy 

\beq
[ a_{u \sigma}, a_{u' \sigma'} ]_+ =0
\;,
\eqlabel{D.10}\eeq

\beq
[ a_{u \sigma}, a_{u' \sigma'}^{\dagger} ]_+ =
\delta_{u,u'}
\delta_{\sigma,\sigma'}
\;,
\eqlabel{D.11}\eeq
where, for any two operators $A$ and $B$, $[A,B]_+ = AB+BA$. 
States in the first quantized formulation are mapped in a 1-1 fashion into
states in the second quantized formulation. For example,

\beq
\ket{+_u}\rarrow a_{u+}^{\dagger}\ket{0}
\;,
\eqlabel{D.12}\eeq

\beq
\ket{-_u}\rarrow a_{u-}^{\dagger}\ket{0}
\;,
\eqlabel{D.13}\eeq

\beq
\ket{\psi^{ant}}
\rarrow
a_{u+}^{\dagger}a_{u-}^{\dagger}\ket{0}
\;.
\eqlabel{D.14}\eeq

From Eqs.(D.4),  (D.12) and (D.13), it follows that 

\beq
\left(
\begin {array}{c}
a_{u+}^{\dagger}\\
a_{u-}^{\dagger}
\end{array}
\right)=
\left(
\begin{array}{cc}
C E^*&SE\\
-SE^*&CE
\end{array}
\right)
\left (
\begin {array}{c}
a_{z+}^{\dagger}\\
a_{z-}^{\dagger}
\end{array}
\right)
\;.
\eqlabel{D.15}\eeq
(To check this last equation, just apply $\ket{0}$ from the right to both
sides of the equation.)

\newpage
\section*{Figure  CAPTIONS:}
\begin{description}

\item{Fig.1}
Node labelled by the random variable $\ul{x}_j$.
\item{Fig.2}
Possible CB nets with 2 nodes $\ul{x}$ and  $\ul{y}$.
\item{Fig.3}
Possible CB nets with 3 nodes $\ul{x}$, $\ul{y}$ and $\ul{z}$.
\item{Fig.4}
Three-node cycle.
\item{Fig.5}
Proof that the joint probability represented by Fig.3d adds up to one
(i.e., satisfies Eq.(2.4) ). Summation over the states of an arrow is
indicated by giving the arrow a double shaft.
\item{Fig.6}
Figure 6a shows the fully connected four node
graph with its chronological labelling. 
By deforming Fig.6a into a topologically equivalent diagram, one
obtains  a more ``stylized'' version of the same thing, Fig.6b.
\item{Fig.7}
Chapman-Kolgomorov equation for a Markov chain
 $(\ul{x},\ul{y},\ul{z})$.
\item{Fig.8}
Generalization of Fig.7 to arbitrary random variables 
$\ul{x},\ul{y},\ul{z}$ that don't necessarily form a Markov chain.
\item{Fig.9}
CB net version of an AND gate.
\item{Fig.10}
$\ul{z}$ is a SUM node.
\item{Fig.11}
$\ul{z}$ is an IF-THEN node. 
\item{Fig.12}
CB net for the Clauser-Horne experiment. 
This figure  really represents 4 nets, one for
each of the following possibilities:  
$(\theta_1, \theta_2) =(A,B), (A,B'), (A',B), (A',B')$.
\item{Fig.13}
CB net  for an experiment like the usual Clauser-Horne 
experiment, except that here  
the measurement
types $\theta_1$ and $\theta_2$ are selected at random.
\item{Fig.14}
CB nets associated with a  random walk.  Net Fig.14c 
is coarser than net Fig.14b, which is coarser than net Fig.14a.
\item{Fig.15}
Root node representing initial wavefunction.
\item{Fig.16a}
Marginalizer node.
\item{Fig.16b}
Phase Shifter node.
\item{Fig.17} 
Nodes for Stern-Gerlach magnet with (a) one, and (b) two incoming modes.
\item{Fig.18}
QB net for an experiment with 2 Stern-Gerlach magnets.
(triangle=wavefunction,
black-filled circle=marginalizer,
white-filled circle=magnet).
\item{Fig.19}
QB net for an experiment with 2 Stern-Gerlach magnets.
(triangle=wavefunction,
black-filled circle=marginalizer,
white-filled circle=magnet).
\item{Fig.20} 
Various sets of evidence that were considered for 
the nets Figs.18 and 19 with two Stern-Gerlach magnets.
\item{Fig.21}
One-hypothesis probabilities, for the net of Fig.19, for evidence-cases 1,
2, 4, 10 and 12. Columns A  to D refer to the parent CB net and
columns F to I to the QB net.
\item{Fig.22}
Two-hypothesis probabilities,
for the net of Fig.19, for evidence-cases 1, 2, 4, 10  and 12. 
Columns A to F  refer to the
parent CB net and columns H to M to the QB net.
\item{Fig.23}
QB net for an experiment with 3 Stern-Gerlach magnets.
(triangle=wavefunction,
black-filled circle=marginalizer,
white-filled circle=magnet).
\item{Fig.24}
QB net for an experiment with 3 Stern-Gerlach magnets.
(triangle=wavefunction,
black-filled circle=marginalizer,
white-filled circle=magnet).
\item{Fig.25}
QB net for an experiment with 3 Stern-Gerlach magnets.
(triangle=wavefunction,
black-filled circle=marginalizer,
white-filled circle=magnet).
\item{Fig.26}
QB net for an experiment with 3 Stern-Gerlach magnets.
(triangle=wavefunction,
black-filled circle=marginalizer,
white-filled circle=magnet).
\item{Fig.27}
QB net for an experiment with 3 Stern-Gerlach magnets.
(triangle=wavefunction,
black-filled circle=marginalizer,
white-filled circle=magnet).
\item{Fig.28}
QB net for an experiment with 3 Stern-Gerlach magnets.
(triangle=wavefunction,
black-filled circle=marginalizer,
white-filled circle=magnet).
\item{Fig.29}
QB net for an experiment with 3 Stern-Gerlach magnets.
(triangle=wavefunction,
black-filled circle=marginalizer,
white-filled circle=magnet).
\item{Fig.30} 
Various sets of evidence that were considered for 
the nets Figs.23 to 29 with three Stern-Gerlach magnets.
\item{Fig.A.1}
Graphical proof of an identity about the intersection of two 
direct product sets.
\item{Fig.C.1}
``The fabric of spacetime".
QB net that yields
the Feynman path
integral for a single mass particle in an arbitrary
potential. 
\item{Fig.C.2}
The input and output arrows for a single node
$(x_s,t_i)$ of the net Fig.C.1. 
\item{Fig.C.3}
In the net Fig.C.1, the  value  $A[n(x_s,t_i)|  n({\bf 
\cdot},t_{i-1})]$ of the node
 $(x_s,t_i)$.
\item{Fig.C.4}
 A typical evaluation of the net of
Fig.C.1. We have blackened those nodes whose state $\ul{n}(x_s,t_i)$ is
1. Arrows coming out of the blackened nodes are also imagined to lie
in state 1.
 We show only those arrows lying in state 1; all other
arrows, i.e., those lying in state 0, are not shown. The value of each
node (the infinitesimal complex amplitude
$\alpha_{s,r}$ for the blackened nodes and approximately 1 for the
unblackened ones)
has been placed next to it.
\item{Fig.D.1}
Definition of the angles
$\theta$ and $\phi$ used in specifying the unit vector $\hat{u}$.

\end{description}


\begin{thebibliography}{99}
\bibitem{pf source}
The phase factor $e^{i \xi}$ does not appear to be a consequence of
the azimuthal phase factors $e^{\frac{\phi}{2}}$ of Eq.(D.5): we've
assumed  the magnetic field directions $\hat{u}$, $\hat{v}$ and $\hat{z}$
to be coplanar, and therefore we may assume that
$\phi_u=\phi_v=\phi_z=0$. 
Suppose $\hat{a},\hat{b},\hat{c}$ are
the particle travel directions  inside 
the magnets with magnetic fields in the directions  
$\hat{u},\hat{v},\hat{z}$, respectively. 
($\hat{a}\cdot\hat{u}=\hat{b}\cdot\hat{v}=\hat{c}\cdot\hat{u}=0$).
The phase factor
$e^{i \xi}$ is still necessary even if
$\hat{a},\hat{b},\hat{c}$
are assumed to be 
parallel, which is a stronger  assumption than only assuming that 
 $\hat{u}$, $\hat{v}$ and $\hat{z}$ are coplanar.
\bibitem{V}It would have been 
just as good an approximation  to use
$V(x_r,t)$ instead of $V(x_s,t)$ in Eq.(C.11). For small $\Delta t$, if 
the kinetic energy term  $\frac{1}{2m}(\frac{x_s-x_r}{\Delta t})^2$ is to
contribute a finite phase to $\alpha_{s,r}$, $x_s$ and $x_r$ must remain
close, and therefore $V(x_r,t)\approx V(x_s,t)$.
\end{thebibliography}
\end{document}